\documentclass[12pt]{article}
\usepackage{cite}
\usepackage{enumerate}
\usepackage{ifpdf}
\usepackage{ae}
\usepackage[T1]{fontenc}
\usepackage[ansinew]{inputenc}
\usepackage{amsmath}
\usepackage{relsize}
\usepackage{amssymb}
\usepackage{tikz}
\usepackage{graphicx}
\usepackage{color}
\definecolor{darkblue}{cmyk}{0.9,0.9,0,0}
\definecolor{darkgreen}{rgb}{0,0.55,0}
\usepackage[colorlinks=true,linkcolor=darkblue,citecolor=darkblue,urlcolor=darkblue]{hyperref}
\usepackage{epsfig}
\usepackage{epstopdf}
\usepackage{graphicx}

\makeatletter
\long\def\@makecaption#1#2{
  \vskip\abovecaptionskip
  \sbox\@tempboxa{{\captionfonts #1: #2}}
  \ifdim \wd\@tempboxa >\hsize
    {\captionfonts #1: #2\par}
  \else
    \hbox to\hsize{\hfil\box\@tempboxa\hfil}
  \fi
  \vskip\belowcaptionskip}
\makeatother

\newcommand{\beq}{\begin{equation}}
\newcommand{\eeq}{\end{equation}}
\newcommand{\beqy} {\begin{eqnarray}}
\newcommand{\eeqy} {\end{eqnarray}}
\newcommand{\bsmat}{\begin{smallmatrix}}
\newcommand{\esmat}{\end{smallmatrix}}
\newcommand{\bmat}{\begin{matrix}}
\newcommand{\emat}{\end{matrix}}

\def\({\left(}
\def\){\right)}
\def\[{\left[}
\def\]{\right]}

\def\<{\langle}
\def\>{\rangle}

\usepackage{varioref}
\usepackage{makeidx}
\makeindex

\usepackage[english]{babel}
\usepackage{caption}
\usepackage{subfig}

\usepackage{tabularx}
\begin{document}

\thispagestyle{empty}

\renewcommand{\thefootnote}{\fnsymbol{footnote}}
\setcounter{page}{1}
\setcounter{footnote}{0}
\setcounter{figure}{0}

\begin{titlepage}

\begin{center}

\vskip 2.3 cm 

\vskip 5mm

{\Large \bf An Algebraic Approach to the Analytic Bootstrap}
\vskip 0.5cm

\vskip 15mm

\centerline{Luis F. Alday$^{z}$ and Alexander Zhiboedov$^{\bar z}$  }
\bigskip
\centerline{\it $^{z}$ Mathematical Institute, University of Oxford,} 
\centerline{\it  Andrew Wiles Building, Radcliffe Observatory Quarter,}
\centerline{\it Woodstock Road, Oxford, OX2 6GG, UK}
\centerline{\it $^{\bar z}$ Center for the Fundamental Laws of Nature,}
\centerline{\it Harvard University, Cambridge, MA 02138 USA}

\end{center}

\vskip 2 cm

\begin{abstract}
\noindent We develop an algebraic approach to the analytic bootstrap in CFTs. By acting with the Casimir operator on the crossing equation we map the problem of doing large spin sums to any desired order to the problem of solving a set of recursion relations. We compute corrections to the anomalous dimension of large spin operators due to the exchange of a primary and its descendants in the crossed channel and show that this leads to a Borel-summable expansion. 
We analyse higher order corrections to the microscopic CFT data in the direct channel and its matching to infinite towers of operators in the crossed channel.
We apply this method to the critical $O(N)$ model. At large $N$ we reproduce the first few terms in the large spin expansion of the known two-loop anomalous dimensions
of higher spin currents in the traceless symmetric representation of $O(N)$ and make further predictions. At small $N$ we present the results for the truncated large spin expansion series of anomalous dimensions of higher spin currents.\end{abstract}

\end{titlepage}

\setcounter{page}{1}
\renewcommand{\thefootnote}{\arabic{footnote}}
\setcounter{footnote}{0}

\newpage

 \def\nref#1{{(\ref{#1})}}
 
 \tableofcontents

\section{Introduction}

Conformal bootstrap uses associativity of the operator algebra to derive constraints on the spectrum of conformal field theories  \cite{Ferrara:1973yt,Polyakov:1974gs}.  In the analytic approach one uses knowledge about the spectrum of light operators in the crossed channel to make statements about certain class of large spin operators in the direct channel. This is the method that we are developing further in the present paper.

Consider a four-point function of scalar operators ${\cal O}$ with dimension $\Delta$. In previous works the leading correction to the anomalous dimensions of the double trace-like operators, schematically ${\cal O}\partial^s {\cal O}$, with scaling dimensions $\Delta_s = 2 \Delta + s + \gamma_s$ was studied   \cite{Alday:2007mf,Fitzpatrick:2012yx,Komargodski:2012ek}. In this paper we present 
an efficient method to take into account sub-leading corrections. These include corrections due to 
\begin{itemize}
\item descendants of a given primary operator;
\item corrections which are of higher order in $\gamma_s$.
\end{itemize}

The procedure we adopt consists of two steps. First, following \cite{Alday:2015eya,Basso:2006nk}, we change variables from the usual spin $s$ to the conformal spin $J$ 
\begin{equation}
\label{conformalspin}
J^2 = (\Delta+s +\gamma_s/2)(\Delta+s +\gamma_s/2-1) .
\end{equation}
This simplifies action of the Casimir operator \cite{Dolan:2000ut} on the conformal blocks. After this change of variables the crossing equation in the limit $v \ll u \ll 1$ becomes
\begin{equation}
\label{crossingequation}
\hat K^{\Delta} \left[ \hat a(J) \left( {u \over 1 - v} \right)^{{\gamma_J \over 2}}  \right] \equiv \int d j \ K^{\Delta}(j,v) \hat a(J)  \left( {u \over 1 - v} \right)^{{\gamma_J \over 2}} = f(v,u),~~~ J^2 = {j^2 \over v} ,
\end{equation}
where $ \hat a(J)$ and $\gamma(J)$ have an expansion in inverse powers of $J$ and $K^{\Delta}(j,v)$ is an integral kernel that can be computed order by order in integer powers of $v$ and comes from the collinear conformal blocks and three-point functions of generalized free fields.

To solve the crossing equations in the limit $v \ll u \ll 1$ amounts to rewrite a given light-cone OPE expansion of $f(v,u)$ in terms of $ \hat a(J)$ and $\gamma_J$ using the known kernel $K^{\Delta}(j,v) $. This is the problem we are solving in the bulk of the paper. It is important to keep in mind that not all the operators in the OPE of $f(v,u)$ in the RHS are dual to the double trace-like operators on the left. It is definitely true for isolated primary operators. But for infinite families of operators with approximately equal twists the dependence on $u$ can significantly depart from the dependence of a single conformal block, schematically of the form $a+ b \log u$. This was discussed in the context of weakly coupled CFTs in \cite{Alday:2015ota}, but the idea is completely general. At present, to understand if a given family of infinite number of operators with almost degenerate twists that appears in the expansion of $f(v,u)$ contributes to the LHS requires a separate discussion and apart from very specific examples, see e.g. \cite{Fitzpatrick:2015qma}, does not seem to be universal.

Second, notice that for all practical purposes it is enough to compute the convolution $\hat K^{\Delta} \left[ J^{- \delta}  \right] $. We find that instead of computing the kernel explicitly and doing the large spin integrals we can act with the Casimir operator ${\cal C}$ on the crossing equation. An immediate consequence of this is an extension of the validity region of the formulas of  \cite{Fitzpatrick:2012yx,Komargodski:2012ek} from  $\Delta - {\Delta_{\epsilon} \over 2} > 0$ to arbitrary value, see \cite{Alday:2015ota}. We also get the following relations
\begin{eqnarray}
\label{casimir}
\hat K^{\Delta} \left[ 1 \right]&=& 1 \ , \\ \nonumber
{\cal C} \hat K^{\Delta} \left[ J^{- \delta} \right]&=& \hat K^{\Delta} \left[ J^{2 - \delta} \right] ,
\end{eqnarray}
where the first equation is the statement that the unit operator in the crossed channel is correctly reproduced. Using some basic properties of the kernel $\hat K^{\Delta}$ and the explicit form of the Casimir operator the relations (\ref{casimir}) can be solved recursively for $\delta= {\rm integer}$ and then analytically continued to arbitrary values of $\delta$.
This method allows one to effectively compute corrections to $\hat a(J)$ and $\gamma_{J}$ without much effort.

Applying this method to the problem that involves a single conformal block in the RHS of (\ref{crossingequation}) we find that the large spin expansions of $\hat a(J)$ and $\gamma_{J}$ are asymptotic, but Borel-summable. This is expected, since, when convoluted with the kernel $\hat K^{\Delta}$, the expansions lead unambiguously to the OPE expansion of $f(v,u)$. We expect this property of the large spin expansions to be completely universal and true in any CFT.  We demonstrate these properties by finding an explicit solution to the problem (\ref{crossingequation}) when on the r.h.s we have a single scalar primary of dimension $2 \times {\rm integer}$. We also briefly describe a toy model that captures qualitative features of the large spin expansion.

Expanding the crossing equation (\ref{crossingequation}) to higher orders in $\left( \gamma_J \right)^k$ we generate terms $(\log u)^k$, with $k \geq 2$, which cannot be reproduced by the exchange of a finite number of operators in the crossed channel. Hence, an infinite number of operators with a particular twist and three-point couplings at large spin is needed in the crossed channel. This was first pointed out in \cite{Fitzpatrick:2015qma}. This leads to accumulation points in the crossed channel. We add two additional observations to this discussion. First, having an accumulation in the crossed channel does not necessarily lead to an enhanced contribution $(\log u)^k$. The universal example of this type is the exchange of the double trace-like operators  themselves. It is easy to see that in this case the enhancement does not occur and we find that the application of the usual formulas of \cite{Fitzpatrick:2012yx,Komargodski:2012ek}, and their generalisations given in the present paper,  to this infinite family of operators leads to the correct result. Moreover, the low spin double trace-like operators in the crossed channel dominate the sum. Second, when the enhancement does occur we find that generically a contribution to the $\log u$ piace and thus $\gamma_J$ is also generated. This contribution, however, requires a detailed knowledge of the coupling to an infinite family of operators that is being exchanged. Thus, accumulation points in the twist spectrum are necessary to satisfy crossing both in the direct and in the crossed channel. Moreover, their effect generically spoils the simple relation between the large spin data in one channel and the OPE data in the other channel.

After a general discussion which is applicable to an arbitrary CFT we apply this method to the critical $O(N)$ model in $2 < d < 4$. We compare the known result for the two-loop, or ${1 \over N^2}$,  anomalous dimensions of higher spin currents in the symmetric traceless representation of $O(N)$, given in \cite{Derkachov:1997ch}, with the prediction of the bootstrap methods. It requires a careful treatment of descendants, accumulation points and higher order corrections in $\gamma_J$ described above. After taking into account all corrections we find a perfect agreement with the result of \cite{Derkachov:1997ch}. We proceed by making a prediction for the two-loop anomalous dimension of the currents in the antisymmetric representation of the $O(N)$ and again find a perfect agreement with the known results of the $4- \epsilon$ expansion\cite{Derkachov:1997ch}. We end up with some all-loop relations between certain terms in the large spin expansion of anomalous dimensions of currents in different representations of the $O(N)$ symmetry.

Finally, we apply the same methods to the case of the $O(N)$ model at small $N$. These models are accessible both experimentally \cite{Pelissetto:2000ek} and using the numerical bootstrap methods \cite{Rattazzi:2008pe}. The new question that arises in this context is how well the large spin expansion approximates the actual result for a given spin $s$. We find that the corrections become small already for $s=4$. This makes us believe that the first few terms in the large spin expansion approximate the actual result with a precision better than $1 \% $. We hope that we can check this statement using the results obtained by other methods in the near future \cite{Kos:2013tga,Kos:2015mba}.

The organization of this paper is as follows. In section two we develop an algebraic approach to perform the sum over spins of double trace-like operators in the direct channel. In section three we apply this approach to the computation of corrections to the anomalous dimensions of higher spin operators due to the exchange of a primary plus all its tower of descendants. In section four we summarize the general picture. In sections five and six we apply this general picture to the critical $O(N)$ model. Finally, we end up with some conclusions and open problems, while many technical details are referred to the appendices.

\section{Anomalous dimensions of higher spin operators: an algebraic approach}

In this section we develop an algebraic approach to the crossing equation in the light-cone limit. In the direct
channel we focus on the contribution of certain higher spin operators. At very large spin these operators 
behave like generalized free fields. To compute corrections to the generalized free field behavior
we go from the sum over spins to the integral. Conformal blocks and three-point functions combine into
a kernel $K^{\Delta}$ which convoluted with the direct channel data should reproduce the crossed channel. The point 
of this section is to show that for practical purposes neither computing the kernel $K^{\Delta}$, nor doing the large spin integral explicitly is actually necessary. Instead it is enough
to understand the effect of acting with the Casimir operator on both sides of the crossing equation. This makes the problem algebraic and allows to solve it much more easily.

\subsection{The sum rule}

Consider a four-point correlator of identical scalar operators ${\cal O}$ of dimension $\Delta$ in a generic CFT.\footnote{We use the following conventions
\begin{eqnarray}
\langle {\cal O} {\cal O} {\cal O} {\cal O} \rangle &=&  {g(u,v) \over (x_{12}^2 x_{34}^2)^{\Delta} }  \ , \\
u &=& z \bar z = {x_{12}^2 x_{34}^2 \over x_{13}^2 x_{24}^2} ,~~~ v=(1-z)(1- \bar z) = {x_{14}^2 x_{23}^2 \over x_{13}^2 x_{24}^2} \ .
\end{eqnarray}} Crossing symmetry implies the crossing equations
\begin{equation}
\label{bootstrap}
v^\Delta \left( 1+ \sum_{\tau,\ell} a_{\tau,\ell} u^{\tau/2} f_{\tau,\ell}(u,v) \right)= u^\Delta \left( 1+ \sum_{\tau,\ell} a_{\tau,\ell} v^{\tau/2} f_{\tau,\ell}(v,u)  \right) \ ,
\end{equation}
where we have singled out the contribution from the identity operator and have written the conformal blocks as $g_{\tau,\ell}(u,v)=u^{\tau/2}f_{\tau,\ell}(u,v)$ to emphasize their leading behavior at small $u$. We will refer to the LHS of (\ref{bootstrap}) as the {\it direct channel} and to the RHS as the {\it crossed channel}. 

As explained in \cite{Fitzpatrick:2012yx,Komargodski:2012ek} the presence of the identity operator in the crossed channel implies the existence of double trace-like higher spin operators, which we denote as $[{\cal O}, {\cal O}]_s$, whose twist approaches $2\Delta$ for large spin. Furthermore, the presence of an operator ${\cal O}_\epsilon$ in the crossed channel leads to a contribution to their anomalous dimension of the form
\begin{equation}
\label{leadings}
\gamma_s =-\frac{c_0}{s^{\tau_\epsilon}} +\cdots
\end{equation}  
where $\Delta_s=2\Delta+s+\gamma_s$ and $c_0$ has been computed in \cite{Fitzpatrick:2012yx,Komargodski:2012ek} . In order to understand this result consider (\ref{bootstrap}) in the small $v$ limit. On the LHS we focus on the contribution arising from the double trace-like operators, with twist $\tau=2\Delta+\gamma_s$. On the RHS we focus on the contribution arising from the operator ${\cal O}_\epsilon$. We obtain the following sum rule
 \begin{equation}
 \label{smallv}
\sum_{s} a_{s} u^{ \gamma_s/2} f_{2 \Delta + \gamma_s, s}(u,v)|_{v \to 0} = a_{\tau_\epsilon} v^{\tau_\epsilon/2 -\Delta} f_{\tau_\epsilon}(v,u)|_{v \to 0}  
\end{equation}
For $\tau_\epsilon/2 -\Delta<0$, the LHS should reproduce a power law divergence near $v =0$. Since each conformal block on the LHS diverges logarithmically as $v \to 0$, this should involve an infinite number of operators, and the divergence will arise from the large $s$ region. In this region the anomalous dimension is very small, and we can consider the piece proportional to $\log u$ in the small $u$ expansion of (\ref{smallv}):
\begin{equation}
\label{sumrule}
\sum_{s} a_{s} u^{ \gamma_s/2} f_{coll}^{(s)}(v) \sim \sum_{s} a_{s}  \frac{\gamma_s}{2} \log u f_{coll}^{(s)}(v)  = a_{\tau_\epsilon} v^{\tau_\epsilon/2 -\Delta} f_{\tau_\epsilon}(v,u)|_{v \ll u \ll 1}   \sim v^{\Delta_\epsilon/2 -\Delta} \log u
\end{equation}
where the collinear conformal blocks are given by 
\begin{equation}
f_{coll}^{(s)}(v)=(1-v)^s ~_2 F_1(\Delta+s+\gamma_s/2,\Delta+s+\gamma_s/2,2\Delta+2s+\gamma_s;1-v).
\end{equation}
The sum over spins on the LHS of (\ref{sumrule}) reproduces the correct divergence provided the behaviour (\ref{leadings}) holds.  More details will be given below. 

\subsection{Algebraic approach}
Let us assume $\Delta$ is sufficiently large and evaluate the divergent contributions coming from the LHS of (\ref{smallv}) as $v \to 0$. As shown in \cite{Alday:2015eya}, the large spin expansion naturally organizes itself in terms of the conformal spin
%
\begin{equation}
J^2 =(\Delta+s +\gamma_s/2)(\Delta+s +\gamma_s/2-1)
\end{equation}
In order to proceed we consider the scaling limit $J^2=j^2/v$, with $v$ very small, and convert the sum over spins into an integral over $j$. Using the integral representation for the hypergeometric function, expanding around $v=0$ and integrating order by order we end up with the following expression (see appendix \ref{CasimirA} for the details)
\begin{equation}
\sum_{s} a_{s} u^{ \gamma_s/2} f_{coll}^{(s)}(v)= v^{-\Delta} \int_0^\infty dj K^\Delta(j,v) \hat a({j^2 \over v}) \left( \frac{u}{1-v}\right)^{\gamma({j^2 \over v})/2} .
\end{equation}
The kernel $K^\Delta(j,v)$ can be computed as an expansion in integer powers of $v$
\begin{eqnarray}
\label{Kgexp}
K^\Delta(j,v) &=& \frac{4}{\Gamma^2(\Delta)} j^{2\Delta-1} K_0(2j) + \cdots.
\end{eqnarray}
For an intermediate operator of twist $\tau_\epsilon$ in the crossed channel, the anomalous dimension and rescaled OPE coefficients (whose precise definition is given in appendix \ref{CasimirA}) have the following expansions
\begin{eqnarray}
\label{gaexp}
\gamma(J)  &=& - {c_0 \over J^{\tau_{\epsilon}}}\left(1+ c_1 \frac{1}{J^2}+\cdots \right) \ , \\ \nonumber
\hat a(J) &=& 1 -{c_0 \over J^{\tau_{\epsilon}}} \left( d_0 + {d_1 \over J^2} + \cdots \right)  \ .
\end{eqnarray}
We could proceed as in \cite{Alday:2015eya}: compute the Kernel $K^{\Delta}(j,v)$ order by order in $v$ and then integrate over $j$ at each order. This will lead to equations from where the coefficients $c_k$ and $d_k$ can be computed, in terms of the RHS of (\ref{smallv}). This however, becomes cumbersome very soon. In the following we will introduce an algebraic approach that will allow us to compute the coefficients $c_k,d_k$ very efficiently.  

First we note that the Kernel $K^\Delta(j,v)$ satisfies the following condition
\begin{equation}
\label{Knorm}
\int_0^\infty dj \ K^\Delta(j,v) =  1 \ , 
\end{equation}
to all orders in $v$. This guarantees we correctly reproduce the contribution from the identity operator in the crossed channel. Next, let us introduce the following family of functions
\begin{equation}
{\cal F}^{(n)}(v) \equiv \int_0^\infty dj \ K^\Delta(j,v) \left( \frac{v}{j^2}\right)^n \ .
\end{equation}
Note that ${\cal F}^{(n)}(v)$ will have a series expansion around $v=0$ that starts at $v^n$. The problem of finding corrections to the anomalous dimension of double trace operators, or OPE coefficients, will then be equivalent to the problem of writing the RHS of (\ref{smallv}), whose details depend on the specific CFT under consideration, in the basis of functions ${\cal F}^{(n)}(v)$.  

\subsection{The Casimir operator and recurrence relations for ${\cal F}^{(n)}(v) $}

In the following we study properties of the ${\cal F}^{(n)}(v)$ functions. Following \cite{Alday:2015eya}, let us start by recalling from \cite{Dolan:2000ut} the existence of a Casimir operator ${\cal C}$ such that

\begin{equation}
\int_0^\infty dj K^\Delta(j,v) h(\frac{v}{j^2}) = F(v) \to \int_0^\infty dj K^\Delta(j,v) \frac{j^2}{v} h(\frac{v}{j^2}) ={\cal C} F(v) \ , 
\end{equation}
where the Casimir operator is given by

\begin{equation}
{\cal C} = \frac{\Delta(\Delta-v)}{v} + (1-v)(1-v-2\Delta)\frac{\partial}{\partial v} + v(1-v)^2 \frac{\partial^2}{\partial v^2} \ .
\end{equation}
From this and the condition (\ref{Knorm}) it follows

\begin{equation}
\label{definition}
{\cal C} {\cal F}^{(n)}(v) ={\cal F}^{(n-1)}(v),~~~~~~ {\cal F}^{(0)}(v) =1  \ .
\end{equation}

Furthermore, as already mentioned, ${\cal F}^{(n)}(v)$ has a series expansion around $v=0$ which starts with $v^n$. These properties can be used to derive recursion relations from which ${\cal F}^{(n)}(v)$ can be efficiently computed order by order in $v$. In order to find the recursion relations it is convenient to write

\begin{equation}
{\cal F}^{(n)}(v)= \frac{v^n}{(1-v)^n} \sum_{\ell=0}^\infty d_\ell^{(n)} \frac{v^\ell}{(1-v)^\ell} \ .
\end{equation}
After some work it is possible to invert (\ref{definition}) to find 

\begin{equation}\label{inverseD}
d_\ell^{(n)} = \sum_{j=0}^\ell \frac{ (-1)^{\ell-j} \, d_{j}^{(n-1)}  }{(j+n-\Delta)(\ell+n-\Delta)} \ ,
\end{equation}
which should be supplemented with $d_\ell^{(0)}=\delta_\ell^0$. These recursion relations can be solved iteratively for $\ell=0,1,2,\cdots$ and any $n$. For instance, for $\ell=0$ we find
\begin{equation}
d_0^{(n)} =\frac{ d_{0}^{(n-1)}  }{(n-\Delta)^2},~~~d_0^{(0)}=1 \ .
\end{equation}
This relation fixes $d_0^{(n)}$ for all integers. For our purposes it will be important to analytically continue the result to non-integer $n$. The easiest way to do this is to compute $d_0^{(n)}$ explicitly from the zeroth order Kernel (\ref{Kgexp}). We obtain

\begin{equation}
d_0^{(n)} =\frac{\Gamma^2(\Delta-n)}{\Gamma^2(\Delta)} \ .
\end{equation}
Which indeed can be seen to satisfy the recursion relations. We stress that this result is valid for generic $n$. Using this expression we can get a recursion relation for $d_1^{(n)}$ and so on. The solutions of these recursions are of the form

\begin{equation}
d_\ell^{(n)} = d_0^{(n)} R_\ell^{(n)} \ ,
\end{equation}
where $R_\ell^{(n)}$ are rational functions. Hence, they can be analytically continued without any ambiguity. For instance, to second order we obtain
\begin{equation}
\label{explicitF}
{\cal F}^{(n)}(v) = \frac{v^n}{(1-v)^n} \frac{\Gamma^2(\Delta-n)}{\Gamma^2(\Delta)}\left(1-\frac{n \left(3 \Delta ^2-6 \Delta +n^2-3 \Delta  n+3 n+2\right)}{3 (-\Delta +n+1)^2} \frac{v}{1-v}+\cdots \right) \ ,
\end{equation}
where $n$ is {\it not} necessarily an integer. Although the resulting expressions become more and more cumbersome, the functions ${\cal F}^{(n)}(v)$ can be computed to any desired order in $v$. 

An important point to keep in mind is that if the small $v$ limit in the crossed channel is $F(v) \sim v^{{\Delta_{\epsilon} \over 2}}$ then after acting with the Casimir operator the small $v$ behavior becomes
$F(v) \sim v^{{\Delta_{\epsilon} - 2 \over 2}}$. Thus, for generic $\Delta_{\epsilon}$ by acting with the Casimir operator a finite number of times we can make it singular. This singularity is reproduced
by the large spin operators in the direct channel. One can see that the result of this matching is equivalent to analytic continuation of formulas  \cite{Fitzpatrick:2012yx,Komargodski:2012ek} away from the $\Delta - {\Delta_{\epsilon} \over 2} > 0$ region (see also appendix A).

\section{Application: contribution from a primary plus all its descendants}

In this section we apply the method developed above to the problem of  exchange of one primary operator in the crossed channel together with its tower of descendants. For generic quantum numbers of the external and exchanged operators we will not find a closed expression for an arbitrary term in the large spin expansion. Instead we present an explicit solution for some particular values of those, including the case of exchange of a scalar primary with dimension 
$\Delta_{\epsilon} = 2 n$ where $n$ is integer. We elucidate general properties of the large spin expansion, showing in particular that it is asymptotic but Borel-summable. 

\subsection{The problem and summary of the results}

The most natural application of the method above is to compute corrections of the form 
\begin{equation}
\label{jexpansion}
\gamma_s =-\frac{c_0}{J^{\tau_\epsilon/2}} \left(1 + \frac{c_1}{J^2}+\frac{c_2}{J^4} +\cdots \right)
\end{equation}  
due to the exchange, in the crossed channel, of an operator ${\cal O}_\epsilon$ plus all its tower of descendants. We will first discuss the case of a scalar operator $\Delta_{\epsilon}$. The conformal block for a scalar operator of dimension $\Delta_\epsilon$ was given in \cite{Dolan:2000ut}
\begin{equation}
f_{\Delta_\epsilon}(v,u)= \sum_{m,n=0} \frac{(\Delta_\epsilon/2)^2_m(\Delta_\epsilon/2)^2_{m+n}}{m! n! (\Delta_\epsilon+1-d/2)_m(\Delta_\epsilon)_{2m+n}} v^m (1-u)^n.
\end{equation}
We are interested in the piece proportional to $\log u$ in the small $u$ expansion. After a short computation we obtain

\begin{equation}
f_{\Delta_\epsilon}(v,u)|_{u \to 0} = -\log u \frac{\Gamma(\Delta_\epsilon)}{\Gamma^2(\frac{\Delta_\epsilon}{2})}~_2F_1\left( \frac{\Delta_\epsilon}{2}, \frac{\Delta_\epsilon}{2},1-\frac{d}{2}+\Delta_\epsilon;v \right) \equiv \log u F_{\Delta_\epsilon}(v)
\end{equation}
This should be reproduced by the corresponding contribution to the anomalous dimension of the double trace operators in the direct channel, namely

\begin{equation}
 \int dj K^\Delta(j,v) \frac{1}{2}\gamma \left(\frac{v}{j^2} \right) = v^{\Delta_\epsilon/2} a_{\Delta_\epsilon} F_{\Delta_\epsilon}(v) \ .
\end{equation}
Plugging the expansion (\ref{jexpansion}) into the sum rule we get
\begin{equation}
-c_0 \left( {\cal F}^{(\Delta_\epsilon/2)}(v)+ \sum_{k=1} c_k {\cal F}^{(\Delta_\epsilon/2+k)}(v)\right) = 2 v^{\Delta_\epsilon/2} a_{\Delta_\epsilon} F_{\Delta_\epsilon}(v) .
\end{equation}
Hence, the problem of finding the coefficients $c_k$ is equivalent to the problem of writing $F_\epsilon(v)$ in the basis of functions $v^{-\Delta_\epsilon/2} {\cal F}^{(\Delta_\epsilon/2+k)}(v)$. Furthermore, since ${\cal F}^{(\Delta_\epsilon/2+k)}(v) \sim v^{\Delta_\epsilon/2+k}$ for small $v$, the system can be truncated to a given order and then expanding in powers of $v$ we obtain a system of equations, from which $c_k$ up to that order can be solved. 

A natural question we would like to answer is what is the nature of the sum 
\begin{equation}
 \hat \gamma(J) \equiv1+\sum_{k=1}^\infty \frac{c_k}{J^{2k}}  \ .
\end{equation}
Does it have a finite radius of convergence, or is it asymptotic? Is it Borel-summable? As we show below, the answer to these questions depend on the precise values of the parameters of the problem, $\Delta$, $\Delta_\epsilon$ and $d$. For isolated cases the series may have finite radius of convergence, but we will find that generically,  the series is asymptotic and Borel-summable. Furthermore, the asymptotic behaviour is universal and we obtain an alternating series with 
\begin{equation}
\left| \frac{c_{k+1}}{c_k} \right| = \frac{k^2}{\pi^2} + \cdots \ ,
\end{equation}
where, provided the series is asymptotic, this leading behaviour is independent of the parameters of the problem. For generic $\Delta_\epsilon$ we present a numerical evidence for this statement, whereas for integer $\Delta_\epsilon/2$ we prove it analytically.

\subsection{Results for general $\Delta_\epsilon$}

We have applied the method above to compute coefficients $c_k$ in the series (\ref{jexpansion}). Such coefficients are complicated functions of the parameters of the problem, namely the dimension $\Delta$ of the external operators, the dimension $\Delta_\epsilon$ of the intermediate primary being exchanged and the spacetime dimension $d$. For instance,
\begin{eqnarray}
c_0 &=&  \frac{2 \Gamma(\Delta_\epsilon)\Gamma^2(\Delta)}{\Gamma^2\left(\frac{\Delta_\epsilon}{2}\right) \Gamma^2\left(\Delta-\frac{\Delta_\epsilon}{2} \right)} a_\epsilon \ , \\ \nonumber
c_1 &=&\frac{\Delta_\epsilon  (-2 d (\Delta_\epsilon  (-3 \Delta +\Delta_\epsilon +3)+2)+\Delta_\epsilon  (-12 (\Delta -1) \Delta +\Delta_\epsilon  (\Delta_\epsilon +4)+8)+8)}{24 (d-2 (\Delta_\epsilon +1))} \ .
\end{eqnarray}  
The leading coefficient $c_0$ agrees with \cite{Fitzpatrick:2012yx,Komargodski:2012ek}. In general the coefficients are of the form
\begin{equation} 
c_k(\Delta,\Delta_\epsilon,d) = \frac{\left( \frac{\Delta_\epsilon}{2}\right)_k}{\left(1- \frac{d}{2}+\Delta_\epsilon \right)_k} P_k(\Delta,\Delta_\epsilon,d) \ ,
\end{equation}
where $(a)_n$ stands for the Pochhammer symbol and $P_k(\Delta,\Delta_\epsilon,d)$ are polynomials of degree $3k$, $2k$ and $k$ in $\Delta_\epsilon$, $\Delta$ and $d$ respectively. Furthermore, they have a parity property: they are even in the variable $\Delta-\frac{d+2}{4}$. For specific values of the parameters, analytic expressions for $c_k$ can be found. First of all, $c_1,c_2,\cdots$ vanish for $\Delta=(d-2)/2$ and $\Delta_\epsilon=2$. As we will see below this is consistent with the results for the critical $O(N)$ model at large $N$. Furthermore, we can find analytic expressions in the following cases
\begin{eqnarray} 
\Delta_\epsilon=1,d=3, \Delta=1/2 ~&\to& ~ c_k = (-1)^{k+1} \frac{(\frac{1}{2})_{k-1}}{2^{2k+1}\Gamma(k+1)} \ , \\ \nonumber
\Delta_\epsilon=1,d=3, \Delta=1 ~&\to& ~c_k= (-1)^{k} \frac{\Gamma(k+\frac{1}{2})}{4^{k}\sqrt{\pi}\Gamma(k+1)} \ .
\end{eqnarray}
In both cases the sum $ \hat \gamma(J) = 1+\sum_{k=1}^\infty \frac{c_k}{J^{2k}}$ has a finite radius of convergence $1/J^2=4$. Finally, let us analyse the expressions for large values of $\Delta$ or $\Delta_{\epsilon}$. For large values of $\Delta$ and $\Delta_{\epsilon} \sim 1$ we obtain
\begin{equation}
P_k(\Delta,\Delta_\epsilon,d) = \frac{\Gamma(k+\frac{\Delta_\epsilon}{2})}{\Gamma(1+k) \Gamma(\frac{\Delta_\epsilon}{2})} \left( \Delta^{2k} - \ell \frac{d+2}{2} \Delta^{2k-1} + \cdots \right) \ .
\end{equation}
In particular, note that we can re-sum the leading terms to obtain
\begin{equation}
\hat \gamma(J) = ~_2F_1(\frac{\Delta_\epsilon}{2},\frac{\Delta_\epsilon}{2},1-\frac{d}{2}+\Delta_\epsilon; \frac{\Delta^2}{J^2}) + \cdots \ .
\end{equation}
This gives the leading contribution in the large $J,\Delta$ limit with $\Delta/J$ kept fixed. For large $\Delta_{\epsilon},\Delta$  and $J$ with $\Delta_{\epsilon}^3 \sim \Delta^3 \sim J^2$ we obtain 
\begin{equation}
\hat \gamma(J)\sim e^{\frac{12 \Delta^2 \Delta_\epsilon+ \Delta_\epsilon^3}{48 J^2}} + \cdots \ .
\end{equation}
For the general case we were not able to find analytic expressions for $c_k$ for arbitrary $k$. However, the method described above can be used to compute a large number of them, and these can be used to study properties of the sum $\hat \gamma(J)$.

\bigskip

\noindent {\bf Example: The Ising model}

\bigskip

An interesting example to study is that of corrections to anomalous dimensions of higher spin operators in the Ising model, due to exchange of the $\epsilon$ operator. In this case we use $d=3$, $\Delta=0.518151$ and $\Delta_\epsilon=1.41264$ \cite{El-Showk:2014dwa}. We have computed  all the coefficients up to $c_{30}$. Figure 1 shows $|\frac{c_{k+1}}{c_k}|$ vs. $k$.

\begin{figure}[h!]
\centering
\includegraphics[width=4in]{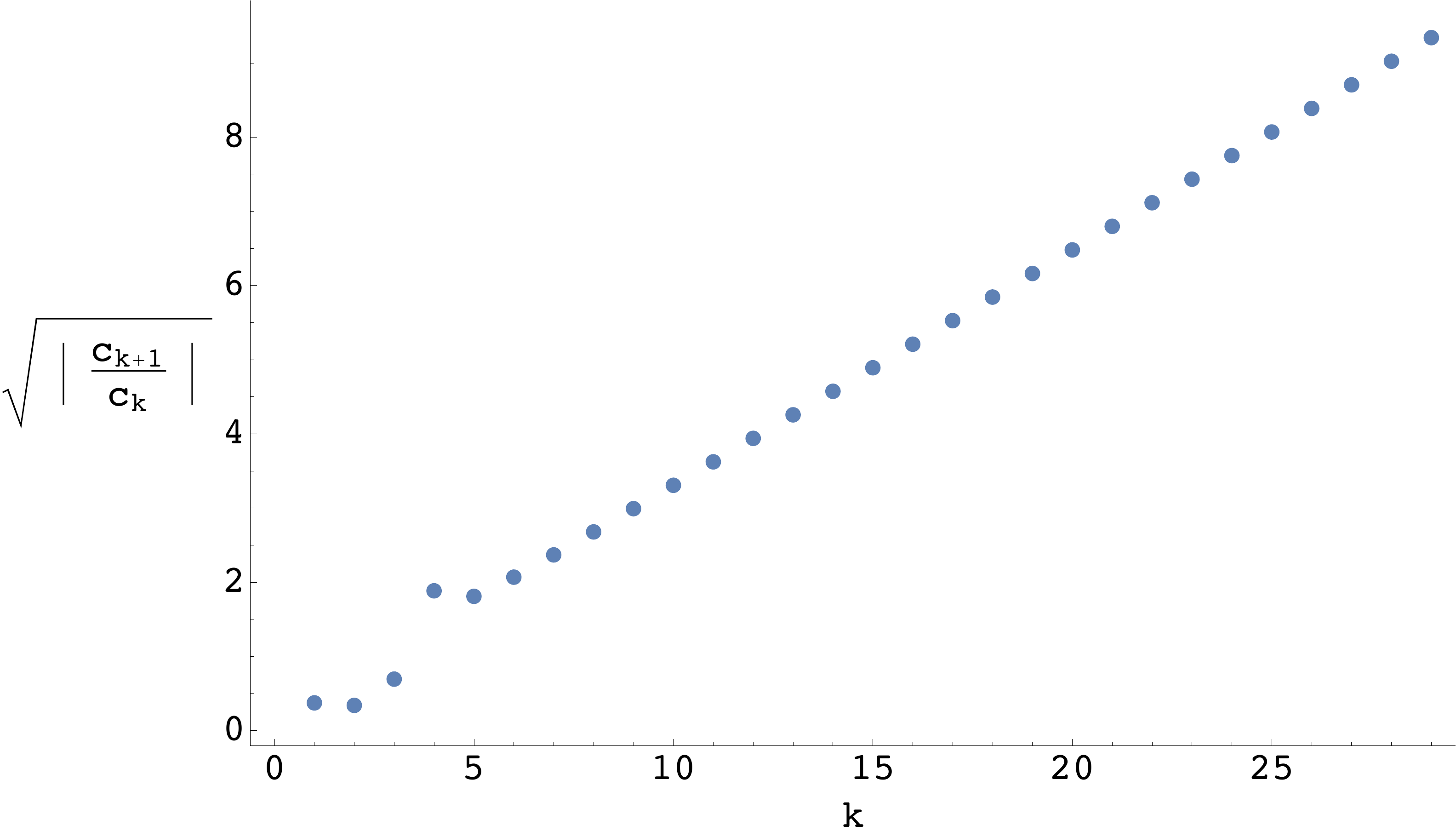}
\caption{Asymptotic behaviour of the coefficients $c_k$ for $d=3$, $\Delta=0.518151$ and $\Delta_\epsilon=1.41264$. The data is consistent with the behaviour $|\frac{c_{k+1}}{c_k}| \sim k^2$ for large $k$. \label{Isingresults}}
\end{figure}

The data is consistent with the following behaviour
\begin{equation}
|\frac{c_{k+1}}{c_k}|= a k^2 \left( 1+ \frac{b}{k} + \cdots \right)
\end{equation}
with $a \sim 0.1$. Furthermore, for large $k$ the series is alternating. This implies that the series is actually asymptotic and Borel-summable. The result for the first few terms is
\begin{eqnarray}
c_1 = 0.06097,~~~~~
c_2 = -0.00846,~~~~~
c_3=0.00097\\ \nonumber
c_4=0.00046,~~~~~
c_5=-0.00165,~~~~~
c_6 =0.00540
\end{eqnarray}
The theory of asymptotic series then suggests  that the best approximation is obtained by including the first few terms (5 or 6, depending of the spin), and the error is very small. For instance, for spin $s=4$, we have $j^2 \sim 12$ and the error is of the order of $c_6/12^6$, which is extremely small.

We have analysed several cases and found that generically the series is Borel summable, with a universal behaviour at large $k$. We will prove this analytically in the next subsection for $\Delta_\epsilon=2m$, with $m$ integer. Before we go ahead, however, let us mention that the asymptotic property and Borel-summability can be understood in terms of a relatively simple toy model, which we discuss in the following. 

\bigskip

\noindent {\bf Toy Model}

\bigskip

Many features of the above computation can be understood with a simple toy model. Let us introduce a toy Kernel such that 

\begin{equation}
\int_0^\infty K^\Delta_{toy}(j,v) \left(\frac{v}{j^2}\right)^n =\frac{v^n}{(1-v)^n}\frac{\Gamma^2(\Delta-n)}{\Gamma^2(\Delta)} \ .
\end{equation}
Now, introducing $\frac{v}{1-v}=\zeta$ 

\begin{equation}
\label{eqtoy}
\int_0^\infty K^\Delta_{toy}(j)  h\left(\frac{v}{j^2}\right) = \sum_{n=0} k_n \zeta^n 
\end{equation}
is solved by 
\begin{equation}
 h(z)= \sum_{n=0} \frac{\Gamma^2(\Delta)}{\Gamma^2(\Delta-n)}k_n z^n \ .
\end{equation}
For the cases at hand, the r.h.s will have a singularity at $v=1$, hence

\begin{equation}
 \sum_{n=0} k_n v^n  \sim \frac{1}{(1-v)^\delta} =  (1+\zeta)^\delta
\end{equation}
Which is a convergent alternating series when expanded around $\zeta=0$ (this is also true for logarithmic singularities). Hence this will lead to a asymptotic expansion, with Borel summable $h(z)$ .  

\subsection{Analytic results for $\Delta_\epsilon=2m$}

In the particular case in which $\Delta_\epsilon/2$ is an integer, we can actually introduce a new formulation which is even more efficient. Let us introduce the variable $\zeta=v/(1-v)$ and define an infinite dimensional vector space such that to each function regular at $\zeta=0$ we associate a vector:
\begin{equation}
f(\zeta) = a_0 + a_ 1 \zeta + a_2 \zeta^2 + \cdots \to \vec{f} =(a_0,a_1,a_2,\cdots)^T \ .
\end{equation}
For instance, ${\cal F}^{(0)}(v) \to \vec{\cal F}^{(0)}=(1,0,0,\cdots)^T$. Now the recurrence relations (\ref{inverseD}) can be written in a very convenient form, namely
\begin{equation}
 \vec{\cal F}^{(n+1)} = {\cal B} \, \vec{\cal F}^{(n)} \ ,
\end{equation}
where ${\cal B}$ is an infinite dimensional matrix whose matrix elements are given by

\begin{eqnarray}\label{matrixB}
{\cal B}_{i,j} = 
\begin{cases}
\frac{(-1)^{i-j+1}}{(i-\Delta)(j-\Delta+1)} ~~~~~i-j \geq 1, \\
~~~~~0~~~~~~~~~~~~~~\textrm{otherwise} \ ,
\end{cases}
\end{eqnarray}
where the indices $i,j$ run from zero. ${\cal B}$ can be interpreted as the inverse of the Casimir operator, although the matrix ${\cal B}$ has zero determinant. 

Let us consider for example the case $\Delta_\epsilon=2$ and let us apply this formalism. In matrix notation the sum rule takes the form

\begin{equation}
c_0 \left(1+ c_1 {\cal B}+ c_2 {\cal B}^{2}+\cdots \right)\vec{\cal F}^{(1)} = \vec F_{\Delta_\epsilon} \ ,
\end{equation}
where remember $\vec{\cal F}^{(1)} ={\cal B} \vec{\cal F}^{(0)}$. The important point is that ${\cal B}^n \vec{\cal F}^{(0)}$ has all its first nth elements equal to zero, so that the above system can be truncated. Furthermore, remember that  $\vec F_{\Delta_\epsilon}$ is given by a hypergeometric function (times a power of $v$). As such, there exist a second order differential operator that annihilates it. In matrix notation
\begin{equation}
{\cal D} \vec F_{\Delta_\epsilon} =0 
\end{equation}
with
\begin{equation}
{\cal D}_{i,j} = (2j-\Delta_\epsilon)(-d+2j+\Delta_\epsilon) \delta_{i,j} +2(2-d+2j)\delta_{i-1,j} \ .
\end{equation}
Hence, we get the following matrix equation for the coefficients $c_k$
\begin{equation}
{\cal D} \left(1+ \sum_\ell c_k {\cal B}^k \right)\vec{\cal F}^{(1)} = 0 \ .
\end{equation}
In practice, we can truncate the matrices to a given order $n \times n$ and then solve for $c_1,\cdots, c_{n-1}$. Just to give an example, for the simplest case $\Delta_\epsilon=2,d=2$ we obtain
\begin{equation}
\label{de2d2}
c_1=\frac{1}{2} (\Delta -2) \Delta,~~~c_2=\frac{1}{3} (\Delta -2)^2 \Delta ^2,~~~c_3=\frac{1}{12} (\Delta -2)^2 \Delta ^2 \left(3 \Delta ^2-6 \Delta -1\right),
\end{equation}
and so on.  Let us stress that this method is very efficient.

%
For general values of $\Delta_{\epsilon}=2m$ the matrix equation becomes
\begin{equation}
{\cal D} \left(1+ \sum_\ell c_k {\cal B}^k \right) {\cal B}^m  \vec{\cal F}^{(0)} = 0 
\end{equation}
from which the coefficients $c_k$ can be computed, as explained above. For $\Delta = 3,4, ...$ they have the general form
\begin{equation}
c_k = \sum_{j=1}^{\Delta-2} b_j(\Delta_\epsilon,\Delta,d) (j(j+1))^k \ ,
\end{equation} 
where $b_j(\Delta_\epsilon,\Delta,d)$ are in general complicated but for any fixed $\Delta_\epsilon=2,4, \cdots$ they can be computed for general $d$ and $\Delta$ (and we have computed them up to $\Delta_{\epsilon}=12$). For instance,
\begin{eqnarray}
b_j(\Delta_\epsilon=2,\Delta,d) =\frac{(d-4) (-1)^j (2 j+1) \Gamma (\Delta -1)^2 \Gamma \left(\frac{d}{2}-\Delta \right) \Gamma \left(-\frac{d}{2}+j+\Delta +1\right)}{2 \Gamma \left(-\frac{d}{2}+\Delta +1\right) \Gamma (-j+\Delta -1) \Gamma (j+\Delta ) \Gamma \left(\frac{d}{2}+j-\Delta +1\right)} \ .
\end{eqnarray}
The limit $d \to \mathcal{Z}$ should be taken with care, as $\Delta$ and $j$ above are also integer.  

Let us next focus in a specific example, lets say $\Delta_\epsilon=2$ and $d=2$. For $\Delta=3,4,\cdots$  we obtain
\begin{eqnarray} 
c_k = \sum_{j=1}^{\Delta-3} \frac{(2j+1)}{(\Delta-1)^2}(j(j+1))^{k} \ .
\end{eqnarray}
Given this, we can compute $\hat \gamma(J) = 1+ \frac{c_1}{J^2} + \cdots$. We obtain

\begin{eqnarray}
\hat \gamma(J)  =  \frac{J^2}{(\Delta-1)^2}\left( \frac{1}{J^2}+  \psi\left( \frac{3}{2}-x \right)+  \psi\left( \frac{3}{2}+x \right)+  \psi\left( \Delta-\frac{1}{2}-x  \right)+  \psi\left(\Delta - \frac{1}{2}+x\right)\right) \ ,
\end{eqnarray}
where we have introduced $x=\frac{1}{2}\sqrt{1+4 J^2}$ and $\psi(z)$ is the digamma function. By using the following identities

\begin{eqnarray}
\psi(-z)&=& \psi(z) + \pi \cot(\pi z) + \frac{1}{z}  \ , \\ \nonumber
\psi(z+1)&=&\psi(z)+\frac{1}{z} \ ,
\end{eqnarray}
we can write
\begin{eqnarray}
\hat \gamma_{an.}(J)  = \frac{2J^2+\Delta+\Delta \sqrt{1+4 J^2}}{2(1-\Delta)(J^2+\Delta-\Delta^2)} + \frac{J^2}{(\Delta-1)^2} \Upsilon(x)  \ ,
\end{eqnarray}
where 
\begin{eqnarray}
\Upsilon(x) =  \psi\left(x+\frac{1}{2} \right)+\psi\left(x-\frac{1}{2} \right)-\psi\left(x+\frac{1}{2}-\Delta \right)-\psi\left(x-\frac{1}{2} +\Delta \right) 
\end{eqnarray}
has been introduced for later convenience. We see that $\hat \gamma(J)$ and $\hat \gamma_{an.}(J)$ differ by
 \begin{eqnarray}
\hat \gamma_{an.}(J)  - \hat \gamma(J)   = \frac{J^2 \pi}{(1-\Delta)^2}\left( \tan(\pi x) - \tan(\pi(x -\Delta)) \right) \ ,
\end{eqnarray}
which vanishes for $\Delta$ integer, so that both expressions agree in this case. On the other hand, for arbitrary $\Delta$, not necessarily integer, $\hat \gamma_{an}(J)$ behaves smoothly for large $J$, whereas $ \hat \gamma(J)$ has an essential singularity. Hence we claim that $\hat \gamma_{an.}(J)$ is the correct analytic continuation, valid for arbitrary $\Delta$. Indeed, expanding it for large $J$ we obtain

\begin{equation}
\hat \gamma_{an.}(J)  = 1 + \frac{\Delta(\Delta-2)}{2 J^2}+ \frac{\Delta^2 (\Delta-2)^2}{3 J^4}+ \frac{(\Delta -2)^2 \Delta ^2 \left(3 \Delta ^2-6 \Delta -1\right)}{12 J^6}+\cdots
\end{equation}
which exactly agrees with (\ref{de2d2}) for any value of $\Delta$ .

We have performed the resummation and analytic continuation for $\Delta_\epsilon=2,4,\cdots$ and $d$ even consistent with the unitary bound. For $d=0,2,\cdots, \Delta_{\epsilon}$ the expressions have always the form
\begin{eqnarray}
\label{analcontinued}
\hat \gamma(J)_{an.}  = {\cal R}_1(x) + {\cal R}_2(x) \Upsilon(x)  \ ,
\end{eqnarray}
where remember $x = \frac{1}{2}\sqrt{1+4J^2}$ and $ {\cal R}_1$ and $ {\cal R}_2$ are rational functions of $x$ which depend of course on $\Delta$ and the specific values of $\Delta_{\epsilon}$ and $d$. Note that for integer values of $\Delta$, $\Upsilon(x)$ becomes a rational function and the expansion in $1/J^2$ will have a finite radius of convergence. On the other hand, as we will show in appendix \ref{asymptoticsA}, for generic $\Delta$ the series is asymptotic, with its asymptotic behaviour controlled by the universal function $\Upsilon(x)$, which is independent of $d$ and $\Delta_{\epsilon}$ .

Finally, let us comment on the re-summed results for $\Delta_\epsilon=2,4,\cdots$ and $d= \Delta_{\epsilon} +2, \Delta_{\epsilon}+4, \cdots,  2\Delta_{\epsilon}$.\footnote{Notice that the scalar operator with dimension $d-2 m$ satisfies unitarity only for $k \leq {d+2 \over 4}$.} In this case, the behaviour is very different and the radius of convergence is finite, even for arbitrary $\Delta$. For instance, for $\Delta_\epsilon=4,d=6$ we obtain
\begin{eqnarray} 
c_k =\frac{1}{2}\left((\Delta-1)((\Delta-1)(\Delta-2))^k -(\Delta-3) ((\Delta-2)(\Delta-3))^k  \right)
\end{eqnarray}
that leads to
\begin{eqnarray}
\hat \gamma(J)_{an.}  =\frac{J^4}{\left(J^2-(\Delta -3) (\Delta -2)\right) \left(J^2-(\Delta -2) (\Delta -1)\right)}
\end{eqnarray}
which has a finite radius of convergence.

\subsection{Exchange of operators with spin}

The large spin expansion of the anomalous dimension of double trace-like operators is controlled by the twist of the exchanged operators. Hence, it is important to take into account exchanged operators with spin. The methods developed in section 2 can be applied in a straightforward way, provided we know enough about the conformal blocks for intermediate operators with spin $v^{\tau_\epsilon/2} f_\epsilon(v,u)$.  For applications to the anomalous dimension, we will be interested in the piece proportional to $u^0 \log u$ in a small $u$ expansion. As we show in appendix C, for an intermediate operator of twist $\tau_\epsilon =\Delta_\epsilon-\ell$ and spin $\ell$ we obtain

\begin{equation}
f_\epsilon(v,u)|_{\log u} = -\frac{(-1)^\ell \Gamma(\Delta_\epsilon+\ell)}{\Gamma^2(\frac{\Delta_\epsilon+\ell}{2})} \left(1+ \sum_{i=1} a^{(i)}_{\Delta_\epsilon,\ell} v^i \right) \ .
\end{equation}
It is shown in the appendix how to compute the coefficients $a^{(i)}_{\Delta_\epsilon,\ell}$. Once these coefficients are known we proceed exactly as for the case of a scalar exchange. We obtain

\begin{equation}
\gamma_s  = -\frac{c_0}{J^{\tau_\epsilon}}\left( 1 + \frac{c_1}{J^2} + \cdots \right) \ ,
\end{equation}
with 
\begin{eqnarray} \nonumber
c_0 &=& a_\epsilon \frac{2(-1)^\ell\Gamma(\tau_\epsilon+2\ell)\Gamma^2(\Delta)}{\Gamma^2(\Delta-\frac{\tau_\epsilon}{2}) \Gamma^2(\ell+ \frac{\tau_\epsilon}{2})},\\
c_1&=&-\frac{(-2 \Delta +\tau_\epsilon +2)^2 \left(2 \ell^2 (d+\tau_\epsilon -2)+2 \ell (\tau_\epsilon -1) (d+\tau_\epsilon -2)+(d-4) \tau_\epsilon ^2\right)}{8 (d+2 \ell-4) (d-2 (\ell+\tau_\epsilon +1))}\\ \nonumber
& & -\frac{1}{12} \tau_\epsilon  (\tau_\epsilon  (-3 \Delta +\tau_\epsilon +3)+2) \ , 
\end{eqnarray}
and so on. $c_0$ above agrees with the result obtained in \cite{Fitzpatrick:2012yx,Komargodski:2012ek}.

Finally, we note the coefficients $c_1,c_2,\cdots$  do not depend on the spin $\ell$ for $\tau_{\epsilon} = d - 2$. We expect that this persists for arbitrary $c_i$ (although we have computed just a few of them) so that the problem is thus reduced to the case of a scalar operator being exchanged. Of course, this case is very special since it describes conserved currents. In this case we can re-sum the whole series coming from descendants to get
\begin{eqnarray}
\hat \gamma(J) &=& {J^{d-2} \Gamma({3 \over 2} - \Delta + {1 \over 2} \sqrt{1 + 4 J^2}) \Gamma(\Delta - {d-1 \over 2}  + {1 \over 2} \sqrt{1 + 4 J^2}) \over \Gamma({1 +d \over 2} - \Delta + {1 \over 2}  \sqrt{1 + 4 J^2}) \Gamma( \Delta - {1 \over 2} + {1 \over 2} \sqrt{1 + 4 J^2})}.
\end{eqnarray}

If we plug the zeroth order value in $\gamma_s$ value for $J^2$ in the formula above we get the following simple formula which takes care of contribution of descendants of conserved currents in general number of dimensions
\begin{equation}\label{resultforstress}
\gamma_s  = - c_0 {\Gamma(s+1) \Gamma(2 \Delta + s - {d \over 2}) \over \Gamma(s + {d \over 2}) \Gamma(2 \Delta + s - 1)} = - {c_0 \over s^{d-2}} + ... \ .
\end{equation}


\section{General picture}

As we have seen in the previous section, given an operator of twist $\tau_\epsilon$ in the cross-channel, the anomalous dimension of higher spin double trace operators receives corrections of the form (\ref{jexpansion}), while the same is true for the corresponding OPE coefficients. For a generic CFT then the naive expectation would be 
\begin{eqnarray}
\label{gaexp}
\gamma^{naive}_s &=&- \sum_{\tau_\epsilon,\ell} \frac{c_0(\tau_\epsilon,\ell)}{J^{\tau_\epsilon}} \left(1 + \frac{c_1(\tau_\epsilon,\ell)}{J^2}+\frac{c_2(\tau_\epsilon,\ell)}{J^4} +\cdots \right) \ , \\ \nonumber
\delta \hat a^{naive}_s &=&- \sum_{\tau_\epsilon,\ell} \frac{c_0(\tau_\epsilon,\ell)}{J^{\tau_\epsilon}} \left(d_0(\tau_\epsilon,\ell) + \frac{d_1(\tau_\epsilon,\ell)}{J^2}+\frac{d_2(\tau_\epsilon,\ell)}{J^4} +\cdots \right)  \ ,
\end{eqnarray}  
where the sum runs over the whole spectrum. For isolated operators in the spectrum, the coefficients $c_k,d_k$ are the ones computed above. In any CFT, however, there are accumulation points, where an infinite number of operators is arbitrarily close to specific twists, so that the sum above is at best formal. When considering the contributions to (\ref{gaexp}) from a tower in an accumulation point, two things can occur:

\begin{itemize}
\item The sum over spins for a given twist is convergent: In this case, the naive expectation (\ref{gaexp}) captures the correct answer. This happens, for example, when double trace-like operators themselves get exchanged.
\item The sum over spins for a given twist is divergent: This is signalling an enhancement of the $\log u$ divergence on the RHS of the sum rule (\ref{sumrule}). In this case the contribution to the RHS from the infinite tower of operators exchanged has to be computed carefully. Examples of this phenomenon are discussed in \cite{Fitzpatrick:2015qma,Alday:2015ota}.
\end{itemize}
This second effect is actually necessary for consistency of the crossing relation. Indeed, plugging the expansions (\ref{gaexp}) into the full sum in the direct channel

\begin{equation}
\label{Kfull}
\int_0^\infty dj K^{\Delta}(j,v) (1+\delta \hat a) \left( \frac{u}{1-v} \right)^{\gamma/2}
\end{equation}
we see that at higher orders in  $\gamma$ we produce higher powers of $\log u$, which can only be reproduced from an infinite number of operators on the crossed channel. 

In order to understand the situation in this case, let us consider a simple example. Suppose the spectrum contains two isolated primary operators of twist $\tau_1$ and $\tau_2$ (such that no other primary or descendant has their twist), and let us consider the contributions from (\ref{Kfull}) to order $v^{\tau_1/2+\tau_2/2}$, arising from quadratic corrections. We obtain
\begin{eqnarray}
c_0(\tau_1)  c_0(\tau_2)\left( \log \left(\frac{u}{1-v} \right) \frac{d_0(\tau_1)+d_0(\tau_2)}{2} + \frac{1}{4} \log^2 \left(\frac{u}{1-v} \right)\right)\int_0^\infty dj \ K^{\Delta}(j,v) \left( \frac{\sqrt{v}}{j} \right)^{\tau_1+\tau_2} \ ,
\end{eqnarray}
which at leading order in $v$ is

\begin{equation}
\label{enh}
c_0(\tau_1)  c_0(\tau_2)\left( \log u \frac{d_0(\tau_1)+d_0(\tau_2)}{2} + \frac{1}{4} \log^2 u \right){\cal F}^{(\tau_1/2+\tau_2/2)}(v) \ .
\end{equation}
As shown in  \cite{Fitzpatrick:2012yx,Komargodski:2012ek} $\tau_1+\tau_2$ will be an accumulation point in the CFT spectrum. It follows that we can consider the exchange of an infinite number of operators with twist close to $\tau_1 + \tau_2$ in the crossed channel.  The microscopic data of the CFT should be such that the quadratic logarithm in (\ref{enh}) is reproduced \cite{Fitzpatrick:2015qma}. Furthermore, the anomalous dimension will generically contain a term of the form
\begin{eqnarray}
\label{gsum}
\gamma_s =- \frac{\tilde{c}_0}{J^{\tau_1+\tau_2}}+\cdots  \ .
\end{eqnarray} 
In order to compute $\tilde{c}_0$, we need to consider its contribution to the $\log u$ term, together with the contribution from (\ref{enh}), and match that to the appropriate $\log u$ piece on the crossed channel. We will perform this explicitly for the examples below. 

\section{Example: the critical $O(N)$ model at large $N$}

In this section we consider the critical $O(N)$ model at large $N$. We consider the ${1 \over N^2}$ correction to the anomalous dimension of the higher spin currents in the symmetric traceless representation of the $O(N)$ model. It was computed explicitly in \cite{Derkachov:1997ch}. We consider the large spin expansion of their result and compare it to the predictions from this paper. In particular this involves taking into account corrections due to infinite families of operators in the crossed channel as well as corrections which are quadratic in $\gamma_s$. The two computations perfectly agree.

\subsection{Known results}

Let us consider the critical $O(N)$ model in $d=2\mu$ dimensions, with $1<\mu<2$. Correlators in this theory can be computed perturbatively in the large $N$ expansion. The spectrum includes the fundamental field $\sigma^i$ of dimension $\Delta_\sigma =  \mu-1+\gamma_\sigma $ and the Lagrange multiplier field $\alpha$  of dimension $\Delta_\alpha = 2+\gamma_\alpha$ with
\begin{eqnarray}
\gamma_\sigma &=& \frac{\eta_1}{N} + \cdots \ , \\ \nonumber
\gamma_\alpha&=& -\frac{4\eta_1}{N} \frac{(\mu-1)(2\mu-1)}{2-\mu} + \cdots \ , 
\end{eqnarray}
where $\eta_1=\frac{2 (2-\mu ) \Gamma (2 \mu -2)}{\Gamma (\mu -1)^2 \Gamma (2-\mu ) \Gamma (\mu +1)}$. We also have an infinite number of operators formed out of these and their derivatives. 

In order to apply the methods of the previous sections, we consider the four-point correlator
\begin{equation}
\langle \sigma_i(x_1) \sigma_j(x_2) \sigma_k(x_3) \sigma_l(x_4) \rangle = \frac{g_{ijkl}(u,v)}{x_{12}^{2 \Delta} x_{34}^{2\Delta}} \ ,
\end{equation}
defining $f_{ijkl}(u,v)=v^\Delta g_{ijkl}(u,v)$, crossing symmetry implies 
\begin{equation}
f_{ijkl}(u,v) = f_{kjil}(v,u) \ .
\end{equation}
The above correlator can be conveniently decomposed into representations of the $O(N)$ global symmetry
\begin{eqnarray}
f_{ijkl}(u,v) &=& \delta_{ij} \delta_{kl} f^S(u,v)+ \frac{1}{2}\left( \delta_{ik}\delta_{jl} + \delta_{il}\delta_{jk} - \frac{2}{N}\delta_{ij} \delta_{kl} \right) f^T(u,v)\\
&+& \frac{1}{2}\left( \delta_{ik}\delta_{jl} - \delta_{il}\delta_{jk}  \right) f^A(u,v) \ ,
\end{eqnarray}
where $S,T,A$ denote the singlet, traceless symmetric and anti-symmetric representations. In this way, we can decompose the crossing relations into three different channels:
\begin{eqnarray}
f^T(u,v)+f^A(u,v) &=& f^T(v,u)+f^A(v,u) \ , \nonumber\\
f^S(u,v)+\left(1-\frac{1}{N} \right) f^T(u,v) &=& f^S(v,u)+\left(1-\frac{1}{N} \right) f^T(v,u) \ , \\
2 f^S(u,v) - \left( 1+\frac{2}{N} \right) f^T(u,v)+f^A(u,v) &=&- 2 f^S(v,u) + \left( 1+\frac{2}{N} \right) f^T(v,u)-f^A(u,v) \ . \nonumber
 \end{eqnarray}
In particular, these relations imply
\begin{equation}
\label{crossingsym}
f^T(u,v)= \frac{1}{2} \left( f^T(v,u) + f^A(v,u)\right) + f^S(v,u) - \frac{f^T(v,u)}{N} .
\end{equation}
We will apply our considerations above to this crossing relation. In the direct channel (LHS of (\ref{crossingsym})) we will consider higher spin operators, in the symmetric traceless representation, of the form $\sigma^{(i} \partial_{i_1} \cdots \partial_{i_s} \sigma^{j)}$.
These operators were considered in \cite{Derkachov:1997ch} where their anomalous dimension was computed to order $1/N^2$. More precisely,
\begin{eqnarray}
\Delta_s = s + 2 \gamma_\sigma + \frac{\gamma^{(1)}_s}{N} +\frac{\gamma^{(2)}_s}{N^2}  + \cdots \ .
\end{eqnarray}

The results are given in formulas (5.24) and (5.25) of that paper (upon taking $\eta_1 \rightarrow 2 \eta_1$ and setting $n \to s$). At one loop the result is
\begin{equation}
\gamma^{(1)}_s= -\frac{2\mu(\mu-1)}{(\mu-1+s)(\mu-2+s)} \eta_1 = -\frac{2\mu(\mu-1)}{J_0^2} \eta_1 \ ,
\end{equation}
where we have written the result in terms of the zeroth order conformal spin $J_0^2=(\mu-1+s)(\mu-2+s)$. Note that at this order we could have written the result in terms of the full conformal spin  $J^2=(\mu-1+s+\gamma_s/2 + \gamma_\sigma)(\mu-2+s+\gamma_s/2+ \gamma_\sigma)$. At two loop all the ingredients of the answer can be readily expanded for large spin except for
\begin{equation}
R(s,\mu)= \int_0^1 \int_0^1 d\alpha d\beta \alpha^{\mu-3} \beta^{\mu-3} (1-\alpha-\beta)^s \ ,
\end{equation}
where the spin is even. We see that for large spin the leading contribution comes from the corners $\alpha,\beta \sim 0$ and $\alpha,\beta \sim 1$. The first corner can be analysed by changing variables $\alpha= A/s,\beta =B/s$, while the second by changing variables  $\alpha= 1-A/s,\beta =1-B/s$. As a result, the large $s$ expansion will consist of two kind of terms. We obtain (see also appendix E)
\begin{eqnarray}
R(s,\mu)= &&\Gamma^2(\mu-2)\frac{1}{s^{2\mu-4}} \left(1+\frac{(3-2\mu)(\mu-2)}{s}+\frac{(\mu -2) (\mu -1) (2 \mu -3) (6 \mu -11)}{6 s^2}+\cdots  \right)  \nonumber \\
&&+\frac{1}{s^2}(1+\frac{3-2 \mu }{s}+\frac{(\mu -2) (3 \mu -2)}{s^2} + \cdots) \ .
\end{eqnarray} 
When expanding the full answer, we obtain different kind of contributions
\begin{equation}\label{twoloop}
\gamma^{(2)}_s = I_1+I_2+I_3+I_4\ ,
\end{equation}
where $I_1$ includes a term proportional to $\frac{\log J_0}{J_0^2}$ plus odd powers of $1/J_0$, $I_2$ includes a single term proportional to $1/J_0^2$, $I_3$ includes terms of the form $\frac{1}{{J_0^{2\mu-2+2n}}}$ and $I_4$ even powers of  $1/J_0$ starting with $1/J_0^4$. More precisely
{\scriptsize
\begin{eqnarray*}
I_1&=& \frac{8 (\mu -1)^2 \mu  (2 \mu -1)}{(\mu -2)} \eta_1^2 \frac{\log J_0}{J_0^2}  + 4\eta_1^2 \frac{\mu(\mu-1)}{J_0^3}-\frac{1}{2}\eta_1^2 \frac{(\mu -1) \mu  (8 (\mu -1) \mu -1)}{J_0^5} +\cdots \ , \\
I_2 &=& -\frac{4(\mu -1) \mu  \left(\frac{\mu  \left(\mu ^2-5 \mu +5\right)}{(\mu -2) (\mu -1)}-\left(2 \mu ^2-3 \mu +2\right) (\psi(2-\mu )-\psi(\mu -1)+\psi(2 \mu -2)+\gamma )+2 (\mu -1) (2 \mu -1) \psi(\mu -1)+1\right)}{\mu -2} \frac{\eta_1^2}{J_0^2} \ , \\
I_3&=& -2\frac{\mu^2(1-\mu)^2\Gamma^2(\mu-2)}{J_0^{2\mu-2}} \eta_1^2 \left(1+\frac{(\mu -3) (\mu -2) (\mu -1)}{3 J_0^2}+\frac{(\mu -3) (\mu -2) \mu  (\mu  (5 \mu -21)+19) (\mu -1)}{90 J_0^4} + \cdots \right) \ , \\
I_4& =& \frac{2 (\mu -1) \mu  (\mu  (4 \mu -3)-4)}{3 (\mu -2)} \frac{\eta_1^2}{J_0^4}-\frac{2  (\mu -1)^2 \mu  (\mu  (15 (\mu -5) \mu +94)-2)}{15 (\mu -2)}  \frac{\eta_1^2}{J_0^6}+ \cdots \ .
\end{eqnarray*}
}
written in terms of the zeroth order Casimir $J^2_0=(\mu-1+s)(\mu-2+s)$ and we kept only first several terms in the expansion of the full answer.

\subsection{Microscopic interpretation}

Let us interpret the above expansion in terms of exchanged operators in the crossed channel, RHS of (\ref{crossingsym}). In the formulas of sections two and three, the dimensions of space-time and of the external operators are to be set to $d=2\mu,~\Delta= \mu-1+ \gamma_\sigma$.

\bigskip
\noindent {\bf Exchange of $\alpha$}
\bigskip

Let us first consider the exchange of the scalar auxiliary field $\alpha$ in the crossed channel. This leads to the following contribution
\begin{equation}
I_\alpha=-\frac{c_0}{J^{\Delta_\alpha}} \left( 1+ \frac{c_1}{J^2}+ \frac{c_2}{J^4} + \cdots \right) \ .
\end{equation}
Where the coefficients $c_k$ are those of section three, with $\Delta_\epsilon=2+\gamma_\alpha$ and the relevant three point function can be found in eq. (21) of \cite{Petkou:1995vu}. Expanding to order $1/N^2$ we get
\begin{equation}
c_0= \frac{2\mu(\mu-1)}{N}\eta_1 +\frac{4 \sin (\pi  \mu ) \Gamma (2 \mu -2) \left(\gamma^{(1)}_\alpha  \psi(\mu -2)+\gamma^{(1)}_\alpha +\frac{2 \gamma^{(1)}_\sigma }{\mu -2}+g_1\right)}{\pi  (\mu -2) N^2 \Gamma (\mu -2)^2} + \cdots \ ,
\end{equation}
where $g_1$ can be found in \cite{Petkou:1995vu}. Furthermore,
\begin{eqnarray}
c_1 &=& -\frac{\gamma^{(1)}_\alpha+6 \gamma^{(1)}_\sigma}{6N} + \cdots,~~~c_2 = \frac{\gamma^{(1)}_\alpha}{30 N} + \cdots,~~~c_3 = -\frac{4 \gamma^{(1)}_\alpha}{315 N} + \cdots,
\end{eqnarray}
etc. Finally, recall that the full Casimir $J^2$ contains $\gamma_s$ itself, which can in turn be first expanded in powers of $1/N$ and then around $J=\infty$. Putting all the pieces together we find:
\begin{equation}
I_\alpha = \frac{I^{(1)}_\alpha}{N} +  \frac{I^{(2)}_\alpha}{N^2}+ \cdots,
\end{equation}
with 
\begin{eqnarray} \nonumber
 I^{(1)}_\alpha &=&- \frac{2\mu(\mu-1) \eta_1}{J_0^2}, \\
I^{(2)}_\alpha &=& \frac{8(\mu-1)^2\mu(2\mu-1) \eta_1^2}{(\mu-2)J_0^2} \log J_0 \\ \nonumber
& & + \frac{4 (\mu -1) \mu \eta_1^2}{J_0^3} -\frac{(\mu -1) \mu  (8 (\mu -1) \mu -1)\eta_1^2}{2 J_0^5} + \cdots  \\ \nonumber
& &+\frac{4 (\mu -1) \mu  \left(\left(6 \mu -4 \mu ^2-2\right) \psi(\mu -2) -\frac{\pi ^{3/2}g_1 2^{1-2 \mu } \csc (\pi  \mu ) \Gamma (\mu +1)}{\Gamma \left(\mu -\frac{1}{2}\right)}-4 \mu ^2 +6 \mu -3\right)}{ (\mu -2) J_0^2}  \nonumber \\
& & +\frac{2 (\mu -1) \mu  (\mu  (4 \mu -3)-4)\eta_1^2}{3 (\mu -2) J_0^4 } -\frac{4 \left((\mu -1)^2 \mu  (2 \mu -1)\right) \eta_1^2}{15 (\mu -2) J_0^6} + \cdots \nonumber \ .
\end{eqnarray}
We see that $I^{(1)}_\alpha$ exactly reproduces the one-loop contribution to $\gamma_s^{(1)}$. Then $I^{(2)}_\alpha$ exactly reproduces $I_1$ and $I_2$ plus the leading term in $I_4$ in (\ref{twoloop}), 

\bigskip
\noindent {\bf Exchange of higher spin operators $\sigma \partial^\ell \sigma$}
\bigskip

Next let us consider the tower of higher spin operators of the form $[\sigma^{i} ,\sigma^{j}]_\ell$, with spin $\ell=0,1,2,\cdots$. Recall that the odd spin operators transform in the anti-symmetric representation of the $O(N)$ symmetry, whereas the even spin ones transform in the symmetric traceless representation. In this case the exchanged operators have twist
\begin{equation}
\tau_\ell = 2\mu-2 +2 \gamma_\sigma +\gamma_\ell,~~~~ \gamma_\ell= -\frac{2\mu(\mu-1)}{(\mu-1+\ell)(\mu-2+\ell)} \frac{\eta_1}{N} + {\cal O}(\frac{1}{N^2}) \ .
\end{equation}
So that the coefficients $c_0(\tau_\ell,\ell)$  become
\begin{equation}
 a_\ell \frac{2(-1)^\ell \Gamma(\tau_\ell+2\ell)\Gamma^2(\Delta)}{\Gamma^2(\Delta-\frac{\tau_\ell}{2}) \Gamma^2(\ell+ \frac{\tau_\ell}{2})} =a_\ell \frac{2(-1)^\ell \Gamma^2(\mu-1)\Gamma(2(\mu+\ell-1))}{\Gamma^2(\mu+\ell-1)} \frac{(\gamma^{(1)}_\ell)^2}{N^2} + \cdots \ .
\end{equation}
Hence, at this order, we can approximate the OPE coefficients $a_\ell$ by their mean field theory value. For the symmetric traceless and anti-symmetric representations
\begin{equation}
 a^{(0)}_\ell = 2 \frac{\Gamma^2(\ell+\mu-1)\Gamma(\ell+2\mu-3)}{\Gamma^2(\mu-1)\Gamma(\ell+1)\Gamma(2\ell+2\mu-3)} \ ,
\end{equation}
while the singlet representation does not contribute at this order. At leading order in $1/J$ we obtain
\begin{eqnarray}
I_{[\sigma,\sigma]} &=& -\frac{1}{2} \frac{1}{J^{2\mu-2}} \sum_{\ell=0}^\infty \frac{4 (-1)^\ell (\mu -1)^2 \mu ^2 (2 \ell+2 \mu -3) \Gamma (\ell+2 \mu -3)}{(\ell+\mu -2)^2 (\ell+\mu -1)^2 \Gamma (\ell+1)} \frac{\eta_1^2}{N^2} + \cdots \\ \nonumber
&=&  -\frac{2\mu^2(1-\mu)^2\Gamma^2(\mu-2)}{J^{2\mu-2}} \frac{\eta_1^2}{N^2} + \cdots \ .
\end{eqnarray}
It is worth noting that the sum over spins above converges very quickly for any $\mu$ and already first few terms approximate the exact answer with the precision better than $1 \%$. This is related to the fact that three-point couplings to the higher spin currents are exponentially suppressed. 

Higher orders in $J$ can be computed from our results in section 3, after setting $\Delta=\mu-1+{\cal O}(1/N),~\tau_\epsilon=2\mu-2+{\cal O}(1/N),~d=2\mu$. At leading order in $1/N$ we obtain
\begin{eqnarray}
c_1 =\frac{(\mu-1)(\mu-2)(\mu-3)}{3},~~~
c_2= \frac{\mu(\mu-1)(\mu-2)(\mu-3)(19+\mu(5\mu-21))}{90}, 
\end{eqnarray}
and so on, so that at order $1/N^2$ we obtain
\begin{eqnarray}
I_{[\sigma, \sigma]} =  -\frac{2\mu^2(1-\mu)^2\Gamma^2(\mu-2)}{J^{2\mu-2}}\left( 1+\frac{(\mu-1)(\mu-2)(\mu-3)}{3 J^2} + \right.\\ \nonumber
\left. + \frac{\mu(\mu-1)(\mu-2)(\mu-3)(19+\mu(5\mu-21))}{90 J^4}+\cdots\right) \frac{\eta_1^2} {N^2} \ , 
\end{eqnarray}
in perfect agreement with $I_3$ of the two loop result (\ref{twoloop}).  

\bigskip
\noindent {\bf Exchange of higher spin operators $[\alpha,\alpha]_\ell$}
\bigskip

Next let us consider double trace operators of the form $[\alpha,\alpha]_\ell$, for $\ell=0,2,\cdots$ . For these operators $\tau_\ell = 4 + {\cal O}(1/N)$ so that the coefficients $c_0(\tau_\ell,\ell)$  become
\begin{equation}
c_0(\tau_\ell,\ell)= a_\ell \frac{2 \Gamma (2 \ell+4) \Gamma (\mu -1)^2}{\Gamma (\ell+2)^2 \Gamma (\mu -3)^2} + \cdots  \ .
\end{equation}
In this case the OPE coefficients $a_\ell$ are already of order $1/N^2$ at large $N$, so that we can stop the expansion at this order. They can be computed using the results for the $\langle \alpha \alpha \sigma^i \sigma^j  \rangle$ four-point function, given in appendix F. We obtain
\begin{equation}\label{threealal}
a_\ell = \frac{\Gamma (\ell+1)^2}{4 (\ell+2) \left(\mu ^2-5 \mu +6\right)^2 \Gamma (2 \ell+2)} \frac{c_0^2}{N^2} + {\cal O}(1/N^3),
\end{equation}
where $c_0=2\mu(\mu-1)\eta_1$ should not be confused with $c_0(\tau_\ell,\ell)$. Note that at this order $c_0(\tau_\ell,\ell) \sim \frac{1}{\ell}$ for large $\ell$, so that if we try to proceed as for the operators above, we would get a divergent sum. This is a manifestation of the phenomenon discussed in section four, and is signalling an enhancement in the logarithmic divergence $\log u \to \log^2u$. Note that this enhancement  is actually necessary: since the crossed channel contains the operator $\alpha$, of approximate twist two, quadratic corrections in the anomalous dimension will produce a $\log^2u$ contribution, which should be reproduced by an infinite tower of operators with twist close to four. These are nothing but the $[\alpha,\alpha]_\ell$ operators.  

\bigskip
\noindent {\bf Exchange of $[\alpha,\alpha]_{\ell,n}$ and descendants of $[\alpha,\alpha]_\ell$}
\bigskip

Corrections due to descendants of $[\alpha,\alpha]_\ell$ as well as due to the exchange of the tower $[\alpha,\alpha]_{\ell,n}$ should be treated similarly. For descendants of $[\alpha,\alpha]_\ell$, it is easy to check that the coefficients $c_k(\ell) \sim 1$ for large $\ell$, so that the sum over spins diverges. We have also computed the corresponding OPE coefficients for $[\alpha,\alpha]_{\ell,n}$, with $n=1$, and the sum over spins diverges as well. 

As discussed in section four, one needs to consider the precise infinite sum in the crossed channel (keeping the full $u$ dependence) and match it to the appropriate contributions in the direct channel, taking into account higher order corrections from $\delta \hat a_s$ and $\gamma_s$. We do this calculation for the first non-trival correction in the following section and  find that the expansion (\ref{twoloop}) is precisely reproduced.

\subsection{Higher order corrections}
Consider the one-loop corrections to the anomalous dimensions and OPE coefficients:
\begin{eqnarray}
\gamma^{(1)} = -\frac{c_0}{N}\frac{1}{J^2} +\cdots,~~~\delta \hat a^{(1)}  = \frac{(\mu-3)(\mu-1)}{2 N} \frac{c_0}{ J^4} + \cdots ,
\end{eqnarray}
where $c_0=2\mu(\mu-1)\eta_1$. These will generate the following contributions to order $1/N^2$ in the direct channel:
\begin{eqnarray}
\label{KquadA}
&&\int_0^\infty dj K^{\Delta}(j,v) (1+\delta \hat a^{(1)} ) \left( \frac{u}{1-v} \right)^{\gamma^{(1)} /2} \\
&=& \frac{c_0^2 \log^2 u}{8} {\cal F}^{(2)}(v) +  \frac{c_0^2 \log u}{4} \left( v  {\cal F}^{(2)}(v)  -(\mu-1)(\mu-3) {\cal F}^{(3)}(v) \right) +\cdots  \nonumber\\
&=& \frac{c_0^2 \log^2 u}{8(\mu -3)^2 (\mu -2)^2} v^2+ c_0^2 \frac{  \log u (13 -4 \mu)+ (7-2 \mu ) \log^2 u}{4 (\mu -4)^2 (\mu -3)^2 (\mu -2)^2}v^3 + \cdots \ . \nonumber
\end{eqnarray}

Let us now turn to the crossed channel. The tower of operators $[\alpha,\alpha]$ have twist $\tau_\ell = 4+ {\cal O}(1/N)$ and they contribute as follows
\begin{equation}
S_{[\alpha,\alpha]} = \sum_{\ell} a_\ell v^2 (1-u)^\ell ~_2F_1(2+\ell,2+\ell,4+2\ell;1-u) \ ,
\end{equation}
where we have kept only the leading  piece for small $v$. Plugging (\ref{threealal}) into the sum we obtain
\begin{equation}
S_{[\alpha,\alpha]} =\frac{v^2}{N^2} \frac{16^{\mu -1} \sin ^2(\pi  \mu ) \Gamma \left(\mu -\frac{1}{2}\right)^2}{\pi ^3 (\mu -3)^2 \Gamma (\mu -1)^2} \frac{\log^2 u}{(1-u)^2} \ .
\end{equation}
We see that indeed, the sum has a $\log^2u$ singularity as $u$ approaches zero. Furthermore, its coefficient precisely agrees with the one in (\ref{KquadA}). Furthermore, note that the correct sum $S_{[\alpha,\alpha]}$ has a vanishing $\log u$ piece. This explains why the $1/J^4$ piece in (\ref{twoloop}) was already reproduced by the exchanged operator $\alpha$ and their descendants! Note that in order to reach this conclusion the full sum needs to be evaluated, and not only its leading piece. 

At next order we have both, primary operators $[\alpha,\alpha]_{\ell,n}$ with $n=1$ as well as the first descendant of the operators just considered. Both of them have twist six at large $N$, so that they will contribute to order $v^3$. The sum to be evaluated is
\begin{equation}
S_{[\alpha,\alpha]_{n=1}+[\alpha,\alpha]_{k=1}} = v^3\sum_{\ell =0,2} \left( a_{\sigma,\sigma,[\alpha,\alpha]_{\ell,n=1}} f^{(0)}_{6+\ell,\ell}(u) + a_{\sigma,\sigma,[\alpha,\alpha]_{\ell}}  f^{(1)}_{4+\ell,\ell}(u) \right) \ ,
\end{equation}
where $f^{(0)}_{\Delta,\ell}(u)$ and $f^{(1)}_{\Delta,\ell}(u)$ are given in appendix B, $a_{\sigma,\sigma,[\alpha,\alpha]_\ell}$ are given in (\ref{threealal}) and $a_{\sigma,\sigma,[\alpha,\alpha]_{\ell,n=1}}$ can be computed to leading order
\begin{eqnarray}
a_{\sigma,\sigma,[\alpha,\alpha]_{\ell,n=1}}  &=& -\frac{\sqrt{\pi } 4^{-\ell-3} (\ell+1) \Gamma (\ell+5)}{(\ell+2) (\ell+3)^2 (\mu -4)^2 (\mu -3) (\ell-\mu +5) (\ell+\mu ) \Gamma \left(\ell+\frac{5}{2}\right)} \frac{c_0^2}{N^2} \ .
\end{eqnarray}
The sums are much more involved in this case, but one can proceed as follows: for each integer $\mu$, the coefficient in front of $(1-u)^n$ can be guessed with some effort. Then the sums can be performed,and the expansion around $u=0$ performed. As a result we get
\begin{equation}
\label{ssum}
S_{[\alpha,\alpha]_{n=1}+[\alpha,\alpha]_{k=1}} = \frac{v^3}{N^2} \left(\frac{\pi^2}{12} \frac{2-\mu}{(4-\mu)^2} +   \frac{16-5\mu}{4(4-\mu)^2}  \log u +  \frac{7-2\mu}{4(4-\mu)^2}  \log^2 u \right) \frac{c_0^2}{(\mu-2)^2(\mu-3)^2} + {\cal O}(u) \ .
\end{equation}
The piece proportional to $\log^2u$ exactly agrees with the corresponding contribution in (\ref{KquadA}). In order to match the piece proportional to $\log u$, we need to consider the linear correction at order $1/N^2$ which is not accounted for by the exchange of $\alpha$ plus its descendants. Namely
\begin{equation}
\gamma_s^{(2)} - I_\alpha = 2(\mu-3)(\mu-1)^2 \mu^2 \frac{\eta_1^2}{N^2} \frac{1}{J^6} + \cdots \ ,
\end{equation}
when integrated agains the Kernel this results in
\begin{equation}
-\frac{\eta_1^2}{N^2}v^3 \frac{(\mu-1)^2 \mu^2 }{(\mu -4)^2 (\mu -3) (\mu -2)^2} = -\frac{c_0^2}{N^2}v^3 \frac{ 1}{(\mu -4)^2 (\mu -3) (\mu -2)^2} \ .
\end{equation}
When added to the $v^3 \log u$ contribution from (\ref{KquadA}) we see that this precisely agrees with what is expected from (\ref{ssum})! 

In summary, we have given a microscopic description of the large spin expansion of the anomalous dimension of higher spin currents in the symmetric traceless representation of the $O(N)$ symmetry up to order $1/J^6$.

\subsection{Currents in other representations}

We can solve the crossing equations for currents in other representations as well. We get
\begin{eqnarray} \nonumber
f^T(u,v) &=& \frac{1}{2} \left( f^T(v,u) + f^A(v,u)\right) + f^S(v,u) - \frac{f^T(v,u)}{N}, \\
f^{A}(u,v) &=& {1 \over 2} ( f^{T}(v,u) + f^{A}(v,u)) - f^{S}(v,u) + { f^{T}(v,u) \over N } , \\ \nonumber
f^{S}(u,v) &=& {1 \over 2}   ( f^{T}(v,u) - f^{A}(v,u)) + {1 \over 2 N} ( f^{T}(v,u) + f^{A}(v,u)) + {f^{S}(v,u) \over N} - {f^{T}(v, u) \over N^2} .
\end{eqnarray}
From this we can make certain predictions about anomalous dimensions of currents in different representations. Let us start with the one-loop discussion.
Recall that the three-point function of $\sigma$ fields with higher spin currents in the singlet representation is ${1 \over N}$ suppressed compared to the other representations. The anomalous dimensions can be found for example in \cite{Lang:1993ge}
\begin{eqnarray} \nonumber
\gamma^{T, (1)}_s &=& - { 2 \mu (\mu -1) \over J_0^2} \eta_1,~~~ s - {\rm even}, \\
\gamma^{A, (1)}_s &=& - { 2 \mu (\mu -1) \over J_0^2} \eta_1,~~~ s - {\rm odd}, \\ \nonumber
\gamma^{S, (1)}_s &=& - { 2 \mu (\mu -1) \over J_0^2} \eta_1 - {\Gamma(2 \mu + 1) \over 2 \mu  - 1} {\Gamma \left({1 \over 2} \sqrt{1 + 4 J_0^2} + 5 - 2 \mu \right) \over J_0^2 \Gamma \left({1 \over 2} \sqrt{1 + 4 J_0^2} - 3  + 2 \mu \right) } \eta_1 .
\end{eqnarray}
Notice that the ${1 \over J_0^2}$ term that comes from the contribution of the $\alpha$ in the crossed channel and is the same for all representations. At higher loops we will get also contributions from the operators involving multiple $\alpha$'s and derivatives. We can focus on the contribution of those by setting all the terms of the type ${1 \over J^{n 2 \mu}}$ in the expansion of $\hat a^{I}$ and $\gamma_s^{I}$ to zero. Then for any two representations we will get the following equation
\begin{eqnarray} \label{relationrep}
\int d j \ K^{\Delta}(j,v) \left( (1 + \delta \hat a^{R_i}) \left( {u \over 1 - v} \right)^{{\gamma^{R_i} \over 2}}  - (1 + \delta \hat a^{R_j}) \left( {u \over 1 - v} \right)^{{\gamma^{R_j} \over 2}} \right) |_{{1 \over J_0^{n \mu} } = 0} = 0 .
\end{eqnarray}

The solution of this equation is 
\begin{eqnarray} \label{relationON}
\gamma_s^{T} = \gamma_s^{A} = \gamma_s^{S}, ~~~ {1 \over s^{\mu}} = 0, \\ \nonumber
\delta \hat a_s^{T} = \delta \hat a_s^{A} =N \delta \hat a_s^{S}, ~~~ {1 \over s^{\mu}} = 0.
\end{eqnarray}
By (\ref{relationON}) we mean that if anomalous dimensions and three-point couplings are expanded in ${1 \over N}$ and ${1 \over s}$ and all ${1 \over s^{n \mu}}$ terms are set to zero then the expressions between different representations should agree. These relations should be valid to any order in ${1 \over N}$ expansion. 

The uniqueness follows from the fact that
\begin{eqnarray}\label{zerokern}
\int d j \ K^{\Delta}(j,v) \sum_{n=0}^{\infty} c_n \left({v \over j^2} \right)^n = 0 
\end{eqnarray}
implies $c_n = 0$. We can apply (\ref{zerokern}) to the terms without and with $\log u$ in (\ref{relationrep}).

At two loops we can make another prediction which follows from the same reasoning as above but by noticing that higher spin currents in the singlet representations do not contribute at this order and the fact that the contribution of the non-singlet currents is the same for both symmetric traceless and anti-symmetric representation up to a sign. We get
\begin{eqnarray} \label{relationON2}
\gamma_s^{A,(2)} &=& \gamma^{S,(2)} - 2 I_3 , \\ \nonumber
I_3 &=& -2\frac{\mu^2(1-\mu)^2\Gamma^2(\mu-2)}{J_0^{2}} \eta_1^2 {\Gamma({1 \over 2} \sqrt{1 + 4 J_0^2} + {5 \over 2} - \mu) \over \Gamma({1 \over 2} \sqrt{1 + 4 J_0^2} - {3 \over 2} + \mu)} \ .
\end{eqnarray}

To our knowledge the result for the currents in the asymmetric representations has not been computed to ${1 \over N^2}$ order. Comparing (\ref{relationON2}) to the computation using $\epsilon$-expansion \cite{Braun:2013tva} we find a perfect agreement. 

\section{The critical $O(N)$ model at small $N$}

In this section we apply the techniques of previous sections to the critical $O(N)$ model at small $N$, namely $N=1,2$. The discussion of this section is much less rigorous than the discussion of previous sections since we truncate the crossed channel sum and estimate the effect of this truncation by evaluating the size of the next few dropped terms. On the other hand, the detailed study of previous sections supports the picture that this is a proper way to handle the large spin expansion sums in CFTs.

As usual we use the known low spin crossed channel data to predict the high spin direct channel data. In the case at hand the low spin data being used consists of: 

\begin{itemize}
\item anomalous dimension of the Lagrange multiplier field $\alpha$ and its coupling to the spin field $\lambda_{\sigma \sigma \alpha }^2$;
\item anomalous dimensions of higher spin currents in different representations of the $O(N)$ symmetry with spin $s \leq 2$ and their coupling to the spin field $\lambda_{\sigma \sigma j_s^{I}}^2$.
\end{itemize}

At small $N$ this data is partially available through numerical bootstrap methods. 
The direct channel data that we would like to compute includes anomalous dimensions of higher spin currents with spin $s \geq 4$ and in various representations.

In principle we have the whole tower of higher spin operators being exchanged in the crossed channel. However, as we observed above, in the case of large $N$ the exchange of high spin operators in the crossed channel is highly suppressed. This follows from the fact that the relevant three-point couplings are exponentially suppressed in the free field theory and also from the ${1 \over \Gamma(\Delta - {\tau_s \over 2})^2}$ in the prefactor. We {\it assume} that this happens at small $N$ and for $s \geq 4$ as well and check that it is a self-consistent assumption at the end. For $s \gg 1$ it is not an assumption and is true in any CFT since the relevant three-point functions are the ones of generalized free fields. Of course, introducing the cut-off in twist and spin in the crossed channel and throwing away everything else is an approximation. The point of this section is to note that for the $O(N)$ model and for 3d Ising it works very well with very few terms.

Note that after truncating the crossed channel we have to consider an exchange of a finite number of primary operators. The formalism for that was fully developed in the bulk of the paper, thus, we simply have to plug the relevant numbers for the particular model.

More precisely, we get the following expressions
\begin{eqnarray}
\tau_s^{I} &=& 2 \Delta_{\sigma} - {c_{\alpha} \over s^{\Delta_{\alpha}}} - \sum_{I', s'=0}^2 {\tilde c_{s'}^{I,I'} \over s^{\tau_{s'}^{I'}} } + ... , 
\end{eqnarray}
where the dots include the contribution both from descendants as well as from heavier primaries and $\Delta_{\sigma}$ is the dimension of the spin field as before. We will estimate the contribution of the latter using the technique of this paper and the contribution of the former when the relevant microscopic data is available.\footnote{One can object that our considerations may not smoothly interpolate to the finite values of spin. One can imagine that the low spin currents do not lie on the smooth curve predicted by the large spin expansion. We simply assume that this does not happen. We thank Slava Rychkov for discussion on that point. }

\subsection{$O(2)$ model}

The relevant low-energy data is \cite{Kos:2013tga,Kos:2015mba}  
\begin{eqnarray} \nonumber
\Delta_{\alpha} &=& 1.511, ~ \Delta_{\sigma} = 0.519, ~ \Delta_{0}^{T} = 1.236, ~ \tau_{2}^{T} =~ ? ~ , \\
{c_{J} \over c_{J}^{free}} &=& 0.905, ~ {c_{T} \over c_{T}^{free}} = 0.944, \\ \nonumber
\lambda^2_{\sigma \sigma \alpha} &=&?~, ~ c_{0,2} = {\lambda^2_{\sigma \sigma j_{0,2}^{T}} \over \lambda^2_{\sigma \sigma j_{0,2}^{T}; free } } =~ ? ~ ,
\end{eqnarray}
where in the last line we introduced the ratios of the actual three-point functions to the free-field ones for spin $0$ and $2$ operators in the symmetric traceless representation of $O(N)$.

Four numbers are missing to write down the prediction for the currents. We leave them as free parameters hoping that they can be determined in the near future, for example, using the numerical bootstrap methods. We present explicitly here the formula for the currents in the traceless symmetric representation of the $O(N)$ which is particularly simple
\begin{eqnarray}\label{otwopr}
\tau_s^{T} - 1 &=& 0.038 - {0.134 \lambda^2_{\sigma \sigma \alpha} \over J_0^{1.511}} - {0.007 \lambda^2_{\sigma \sigma \alpha} \over J_0^{3.511}} - {5 \cdot 10^{-5} \over J_0} - {5.5 \cdot 10^{-6} \over J_0^3} , \\ \nonumber
J_0^2 &=& (s + 0.519)(s-0.481) ,
\end{eqnarray}
and depends only on one unknown parameter. The leading spin corrections to the anomalous dimension of currents were recently computed in \cite{Li:2015rfa} and are in a perfect agreement with the formula above.

Formulas for currents in other representations could be easily obtained using the general formulas in the bulk of the paper, see also  \cite{Li:2015rfa}.

\subsection{3d Ising}

The relevant low-energy data is   \cite{ElShowk:2012ht,El-Showk:2014dwa}
\begin{eqnarray} \nonumber
\Delta_{\alpha} &=& 1.4127,~ \Delta_{\phi} = 0.5182, \\
{c_{T} \over c_{T}^{free}} &=& 0.9465, ~ \lambda^2_{\phi \phi \alpha} = 1.1064 .
\end{eqnarray}
Plugging this data into the generic formula for anomalous dimension we get
\begin{eqnarray}
\tau_s - 1 &=& 0.0364 - {0.0925 \over J_0^{1.4127}} - {0.0027 \over J_0} - {0.0056 \over J_0^{3.413}} - {0.0003 \over J_0^3}, \\ \nonumber
J_0^2 &=& (s + 0.5182)(s-0.4818).
\end{eqnarray}
In this formula we kept only the leading correction due to descendants because with the precision we are working we found that the contribution of the higher order corrections is negligible for $s \geq 4$. More precisely, we find that the correction due to other descendants or heavier primaries for $s=4$ is of order $\pm 0.001$. Of course, our estimate is not rigorous because we do not have the full control over the crossed channel sum. Also we kept only the contribution of the stress tensor from the tower of higher spin currents. In our previous work \cite{Alday:2015ota} we left the coefficient ${c_0 \over J_0}$ unknown. As explained above the contribution of higher spin currents to $c_0$ is small compared to the one of the stress tensor.

\section{Conclusions}

In this paper we considered the crossing equations for conformal field theories in the light-cone limit. Based on the previous work \cite{Alday:2015eya} we developed a systematic method
to compute further corrections to the double trace-like operators in the direct channel using the low energy data in the crossed channel, generalising the results of \cite{Fitzpatrick:2012yx,Komargodski:2012ek} to arbitrarily high order. The basic idea is first to switch 
from the usual spin $s$ to the conformal spin $J$ and then analyse the effects of acting on the crossing equation with the Casimir operator. As we showed in section 2 it makes
the problem effectively algebraic. 

We considered the problem of exchange of a single primary together with its descendants in the crossed channel. Our systematic procedure allows to compute arbitrarily high corrections to the anomalous dimension of double trace-like operators for the generic case. For the special case of exchange of a scalar operator whose dimension is an even integer, we found a closed expression for the n'th correction. We found that generically the expansion is asymptotic but Borel summable. This feature can be understood from a relatively simple toy model. Furthermore, the toy model suggests that Borel summability is a feature of the expansion of the full anomalous dimension, and is intimately connected to the fact that the anomalous dimension results in a convergent, unambiguous expression when integrated against the Kernel. We also discussed the general picture of the large spin expansion in a generic CFT that emerges from previous considerations in section 4 and applied it to the critical $O(N)$ model in sections 5.

The main new feature of the critical $O(N)$ example is several infinite families of operators being exchanged in the crossed channel. The contribution of higher spin currents and their descendants is taken into account by simply summing over the results we got for the exchange of a single primary. On the other hand, operators $[\alpha, \alpha]_{n, \ell}$ contribute to the quadratic, as predicted by \cite{Fitzpatrick:2015qma}, as well as to linear in $\gamma_s$ terms. After taking into account all corrections we found perfect agreement with the perturbative result of \cite{Derkachov:1997ch}. Moreover, using crossing equations we predicted a two-loop result for currents in the anti-symmetric representation of the $O(N)$ symmetry. Comparing it with the known $4 - \epsilon$ expansion result \cite{Braun:2013tva} we found perfect agreement. We also made some further all-loop predictions.

At last we applied the large spin methods to the $O(N)$ models with small $N$. The new feature in this case is that we truncate the large spin expansion series and ask how good this approximation is. 
Treating the large spin expansion as an asymptotic series expansion and estimating the error by the size of the first dropped term we found that already for $s=4$ the error is very small. It would be nice to compare this picture with the results obtained by some other methods, most probably the ones of numerical bootstrap.

We hope that the present analysis will help to develop connections between the bootstrap methods and other methods, be it the usual perturbation theory or integrability, initiated in \cite{Alday:2013cwa,Alday:2015ota}. Already in the case of the $O(N)$ model it became clear that bootstrap techniques allow to understand the perturbative results microscopically and lead to further predictions. Conversely having an explicit perturbative result allowed us to test very general ideas about the crossing equation. One may hope that this instance is not a single one and we can learn many more things by combining different techniques.

There are many directions in which the present analysis can be extended. The most immediate one would be to push further the application to the $O(N)$ model, for instance in the $\epsilon$-expansion. One could also extend the methods to the case of external operators with spin. Furthermore, it would be interesting to apply this algebraic approach other conformal field theories, e.g. gauge theories dual to string theories. Perturbative gauge theories, along the lines of  \cite{Alday:2013cwa,Alday:2015ota} or gauge theories with a large central charge, along the lines of \cite{Alday:2014tsa}, would be two very interesting examples. 

Another extension is to generalize the analysis to the case of operators ${\cal O} (\partial^2)^n \partial^{\ell} {\cal O}$ with non-zero $n \neq 0$ \cite{Kaviraj:2015cxa}. Taking into accounts the descendants and considering $n \gg 1$ limit we expect to recover the full AdS propagator along the lines of \cite{Cornalba:2007zb}. This together with \cite{Hartman:2015lfa} may be useful for deriving \cite{Camanho:2014apa} directly from bootstrap. Another open problem is more systematic understanding of convergence properties of the large spin expansion and errors made when truncating the series. This seems to be necessary to connect analytic bootstrap results to the ones of numerical bootstrap \cite{Rattazzi:2008pe}. Our results (and the study of the toy model) suggest that such expansions are asymptotic and Borel-summable, however, since the spectrum of a CFT becomes denser and denser as we increase the twist, it is hard to make rigorous claims. On a more conceptual level it would be interesting to understand if analytic bootstrap can be used to say anything about the landscape of possible CFTs. At present the spectrum in one of the channels is given and then consequences are derived. In the case of weakly coupled CFTs one can do better \cite{Alday:2015ota}, but it is not clear if this approach can be made completely general.

\section*{Acknowledgments}

We are grateful to A. Bissi, S. Giombi and A. Manashov for useful discussions. We thank David Poland for pointing out an error in the original version of the formula (\ref{otwopr}). The work of L.F.A was supported by ERC STG grant 306260. L.F.A. is
a Wolfson Royal Society Research Merit Award holder. A.Z. is supported in part by U.S. Department of Energy grant de-sc0007870.
 
\appendix

\section{The Kernel and the Casimir operator}
\label{CasimirA}

Given the four-point correlator of identical operators ${\cal O}$, the contribution from double trace operators $[{\cal O},{\cal O}]$ is proportional to the following sum:
\begin{equation}
\label{sum}
S=\sum_{s=0,2,\cdots} a_{s} u^{ \gamma_s/2}(1-v)^s ~_2 F_1(\Delta+s+\gamma_s/2,\Delta+s+\gamma_s/2,2\Delta+2s+\gamma_s;1-v) .
\end{equation}
We are interested in evaluating the divergent contributions, as $v \to 0$, from such a sum. In a conformal field theory of generalised free fields, the OPE coefficients take the following value 
\begin{equation}
\label{freeOPE}
a_s=\frac{2 \Gamma \left(\Delta +s\right)^2 \Gamma (2\Delta+s-1)}{\Gamma (s+1) \Gamma \left(\Delta\right)^2 \Gamma (2\Delta+2s-1)} \ .
\end{equation}
One can explicitly check that for small values of $v$, divergent contributions comes from large spins, of order $s \sim \frac{1}{v^{1/2}}$. Based on this, we consider the scaling limit $s=\frac{x}{\sqrt{v}}$ and convert the sum over spins into an integral over $x$:
\begin{equation}
\sum_s \to \frac{1}{2} \int dx \ ,
\end{equation}
where the factor of $1/2$ is due to the fact that only even spins contribute. Furthermore, we consider the following standard integral representation for the hypergeometric function
\begin{equation}
\label{sum}
 ~_2 F_1(a,b,c;z)= \frac{\Gamma(c)}{\Gamma(b)\Gamma(c-b)} \int_0^1 \frac{t^{b-1}(1-t)^{c-b-1}}{(1-t z)^{a}} dt \ ,
\end{equation}
and change variables $t \to \lambda$, with $t=1-\lambda \sqrt{v}$, such that the important contributions will come from a finite range in the $\lambda$ variable.  Finally, it is convenient to perform a final change of variables, from $x \to j$, with
\begin{equation}
\frac{j^2}{v} = \left(\frac{x}{\sqrt{v}}+ \frac{\gamma(\frac{x}{\sqrt{v}})}{2} +\Delta-1 \right) \left(\frac{x}{\sqrt{v}}+ \frac{\gamma(\frac{x}{\sqrt{v}})}{2} +\Delta \right) \ .
\end{equation}
In terms of the redefined structure constants
\begin{equation}
\label{rescaledOPE}
a_s=\frac{2 \Gamma \left(\Delta + s+ \frac{\gamma_s}{2} \right)^2 \Gamma (s+2\Delta+ \frac{\gamma_s}{2} -1)}{\Gamma (s+1+ \frac{\gamma_s}{2}) \Gamma \left(\Delta\right)^2 \Gamma (2 s+2\Delta+\gamma_s -1)}(1+\frac{1}{2} \partial_s \gamma_s) \hat a_s\,.
\end{equation}
we end up with the following expression
\begin{equation}
S= v^{-\Delta} \int_0^\infty dj \ K^{\Delta}(j,v) \hat a(\frac{j^2}{v}) \left(\frac{u}{1-v} \right)^{\gamma(\frac{j^2}{v})/2} \ ,
\end{equation}
where now $\hat a$ and $\gamma$ are interpreted as functions of the conformal spin $J^2=j^2/v$, and the Kernel is given by
\begin{eqnarray}
K^\Delta(j,v) &=& \frac{2 j v^{\Delta-1}}{(1-v)^\Delta} \frac{\Gamma \left(\Delta +\frac{1}{2} \left(\sqrt{\frac{4 j^2}{v}+1}-1\right)\right)}{\Gamma \left(\frac{1}{2}\sqrt{\frac{4 j^2}{v }+1}+\frac{3}{2}- \Delta \right) \Gamma^2(\Delta)} \times \\
& & ~~~~~~~~~~~~~~ \times \int_0^\infty \frac{d\lambda }{\lambda(1-\sqrt{v}\lambda)} \left(\frac{\lambda  (1 - v) \left(1 - \lambda  \sqrt{v}\right)}{\lambda - v \lambda  +\sqrt{v}}\right)^{\frac{1}{2} \left(\sqrt{\frac{4 j^2}{v }+1}+1\right)} \ .
\end{eqnarray}
The Kernel can be explicitly computed to any desired order in $v$, by expanding and then integrating. To leading order in $v$:
\begin{eqnarray}
K^\Delta(j,v) = \frac{4}{\Gamma^2(\Delta)} j^{2\Delta-1} K_0(2j) + \cdots  \ .
\end{eqnarray}
The normalisation property (\ref{Knorm}) can be verified order by order in $v$, to any desired order. The property 
\begin{equation}
\int_0^\infty dj \ K^{\Delta}(j,v) h(\frac{v}{j^2}) = F(v) \to \int_0^\infty dj  \ K^{\Delta}(j,v) \frac{j^2}{v} h(\frac{v}{j^2}) ={\cal C} F(v)
\end{equation}
is easier to check at the level of the original sum, since each collinear conformal block is an eigenfunction of the Casimir operator. 

Let us discuss in detail how the Casimir operator can be used in order study the contributions from higher spin operators to the sum $S$. First, let us make clear the limitations of the method: we will be able to capture only combinations that become singular as $v \to 0$, upon the application of the Casimir operator a finite number of times. In particular, a finite number of terms in the sum over conformal blocks will not produce the required behaviour. A finite number of conformal blocks has only a logarithmic divergence as $v \to 0$, and this will remain so, no matter how many times the Casimir operator is applied. 

In order to see precisely how the method works, let us consider 
\begin{equation}
\label{Kintegral}
\int_0^\infty dj \ K^{\Delta}(j,v) h(\frac{v}{j^2}) = v^\Delta F_{reg}(v)  \ ,
\end{equation}
where $F_{reg}(v) = a_0(\log v)+ v a_1(\log v)+\cdots$ does not contain power-law divergences as $v \to 0$, but can contain logarithmic divergences. The would-be solution $h(v/j^2)$ has then an expansion of the form
\begin{equation}
h(\frac{v}{j^2}) = \left( \frac{v}{j^2}\right)^\Delta h_0(\log \frac{v}{j^2} ) + \cdots \ .
\end{equation}
But upon plugging this into (\ref{Kintegral}) we see that the integrand has a non-integrable singularity in the small $j$ region. This is a direct consequence of $F_{reg}(v) $ in (\ref{Kintegral}) not being divergent enough as $v \to 0$. In order to proceed, let us apply the Casimir operator on both sides:
\begin{equation}
\label{Kcas}
\int_0^\infty dj \left( \frac{j^2}{v} \right) K^{\Delta}(j,v) h(\frac{v}{j^2}) = v^\Delta \left( \frac{a_0''(\log v)}{v} + \cdots  \right) \ .
\end{equation}
On the LHS this has ameliorated the behaviour of the integrand in the small $j$ region. On the RHS this has enhanced the divergence as $v \to 0$. Now we can safely plug $$h(\frac{v}{j^2}) = \left( \frac{v}{j^2}\right)^\Delta h_0(\log \frac{v}{j^2}) + \cdots \ .$$ and obtain an equation for the leading order $h_0(\log  \frac{v}{j^2})$:
\begin{equation}
\label{leadingeq}
\frac{4}{\Gamma^2(\Delta)} \int_0^\infty dj j K_0(2j) h_0(\log\frac{v}{j^2}) = a_0''(\log v) \ .
\end{equation}
Now all integrals involved are perfectly convergent. Note a very important point. In order for this contribution to be non-zero, $a_0(\log v)$ needs to diverge at least as $\log^2 v$ as $v \to 0$, this can only be possible if an infinite number of operators contributes to the sum, in agreement with the discussion above. Finally, let us mention that in most applications  $F_{reg}(v) $ will contain an extra-parameter (such as the twist of exchange operators) so that the sum will actually be of the form
\begin{equation}
\int_0^\infty dj \ K^{\Delta}(j,v) h(\frac{v}{j^2}) = v^{\Delta_\epsilon/2}  \left(a_0(\log v)+ \cdots  \right) \ .
\end{equation}
The solution is now 
$$h(\frac{v}{j^2}) = \left( \frac{v}{j^2}\right)^{\Delta_\epsilon/2} h_0(\log v) + \cdots  $$
with 
\begin{equation}
\frac{4}{\Gamma^2(\Delta)} \int_0^\infty dj j^{2\Delta-1} K_0(2j)  \left( \frac{v}{j^2}\right)^{\Delta_\epsilon/2}  h_0(\log\frac{v}{j^2}) = v^{\Delta_\epsilon/2} a_0(\log v) \ .
\end{equation}
We can solve for $h_0$ in the region $2\Delta>\Delta_\epsilon$ and then analytically continue. Note that to the leading order we are considering, acting with the Casimir operator has the same effect as changing $\Delta_\epsilon \to \Delta_\epsilon - 2$, so that acting with the Casimir operator (for a case in which the integrals do not converge) is equivalent to analliticaly continue in $\Delta_\epsilon$.

\section{Asymptotics of the universal function}
\label{asymptoticsA}

We want to study the asymptotic expansion of $\Upsilon\left(\frac{1}{2}\sqrt{1+4J^2}\right)$, with
\begin{equation}
\Upsilon(x) =  \psi\left(x+\frac{1}{2} \right)+\psi\left(x-\frac{1}{2} \right)-\psi\left(x+\frac{1}{2}-\Delta \right)+\psi\left(x-\frac{1}{2} +\Delta \right) \ .
\end{equation}
At leading order we can approximate $\frac{1}{2}\sqrt{1+4J^2} = J + {\cal O}(1/J)$ so that we are interested in the asymptotic behaviour of
\begin{equation}
\Upsilon_{asym}(J)= \psi\left(J+\frac{1}{2} \right)+\psi\left(J-\frac{1}{2} \right)-\psi\left(J+\frac{1}{2}-\Delta \right)-\psi\left(J-\frac{1}{2} +\Delta \right) 
\end{equation}
the above can be expanded as
\begin{eqnarray}
\Upsilon_{asym}(J)= \sum_{n=1}^\infty \frac{1}{(2n)! 2^{2n-1}} \left( 1 - (2\Delta-1)^{2n} \right) \psi_{2n} \left( J \right) \ .
\end{eqnarray}
The asymptotic expansion of the polygamma function is given by
\begin{equation}
 \psi_{2n} \left( J \right) = -\frac{(2n-1)!(2n+2J)}{2 J^{2n+1}} - \sum_{k=1}^\infty \frac{(2k+2n-1)!}{(2k)! J^{2k+2n}} B_{2k} \ ,
\end{equation}
where $B_{2k}$ are the even Bernoulli numbers. The first term in this expression, as well as any finite number of terms in the sum, will lead to an expression with a finite radius of convergence. Hence the leading asymptotic behaviour will come from the large $k$ region. Rearranging the sums and discarding the first term we obtain
\begin{eqnarray}
\Upsilon_{asym}(J) \sim \sum_{m}^\infty \frac{(2m-1)!}{J^{2m}} \sum_{n=0}^m \frac{(2\Delta-1)^{2}-1}{2^{2n-1}(2n)!(2(m-n)!)} B_{2(m-n)} \ .
\end{eqnarray}
We would like to estimate the sum for large values of $m$. Using the asymptotic expression for the Bernoulli numbers
\begin{eqnarray}
B_{2n} \sim (-1)^{n-1} 4 \sqrt{\pi n} \left(\frac{n}{\pi e} \right)^{2n}
\end{eqnarray}
we obtain
\begin{eqnarray}
\Upsilon_{asym}(J)&=& -\sum_{m}^\infty \frac{4}{J^{2m}}\frac{(-1)^m (2m-1)!}{4^m \pi^{2m}} \sum_{n=0}^\infty \frac{(-1)^n \pi^{2n} ((2\Delta-1)^{2n}-1)}{(2n)!}\\ \nonumber
&=& -\sum_{m}^\infty \frac{4}{J^{2m}}\frac{(-1)^m (2m-1)!}{4^m \pi^{2m}} (1-\cos(2\pi \Delta)) \ .
\end{eqnarray}
For non-integer $\Delta$, we get alternating coefficients growing like $\Gamma(2m)$. From this we can compute the leading asymptotic behaviour
\begin{equation}
|\frac{c_{m+1}}{c_m}| = \frac{m^2}{\pi^2}  + \cdots \ .
\end{equation}
Note that the presence of overall rational functions will not change this behaviour, so this result should be universal. In particular, note that this perfectly agrees with the numerical results for the 3d Ising model, see section 3.2.

\section{Conformal blocks with spin}

In this appendix we compute conformal blocks for intermediate operators with spin, in a particular limit which we describe bellow. We are interested in the case were the external operators are all identical scalars. In this case, \cite{ElShowk:2012ht} gave sets of recursion relations to compute the conformal blocks for a given spin $G_{\Delta,\ell}(u,v)$ in terms of lower spin conformal blocks. The relation between $G_{\Delta,\ell}(u,v)$ and the conformal blocks used in this paper is given by:
\begin{equation}
u^{(\Delta-\ell)/2} f_{\Delta,\ell}(u,v)= (-1)^\ell \frac{(d-2)_\ell}{(d/2-1)_\ell} G_{\Delta,\ell}(u,v)  \ .
\end{equation} 
Among the recurrence relations given by \cite{ElShowk:2012ht}, the one we found most useful is
\begin{eqnarray}\nonumber
& &\frac{(\Delta-\alpha)(\ell+2\alpha-1)}{\ell+\alpha-1} G_{\Delta,\ell} = \left(\frac{1}{2}(\Delta+\ell-2\alpha-2)\frac{1-v}{u}-2 v \partial_v +(1-u-v) \partial_u \right) G_{\Delta+1,\ell-1}\\
&-&(\ell-1)\left( \frac{\Delta(\Delta-2\alpha+1)}{(\Delta-\alpha)(\Delta-\alpha+1)} \beta_{\frac{1}{2}(\Delta-\ell+2-2\alpha)} G_{\Delta+2,\ell-2} + \frac{\Delta+\ell-1}{\ell+\alpha-1} G_{\Delta,\ell-2} \right) \ ,
\end{eqnarray} 
where $\alpha=d/2-1$ and $\beta_p=\frac{p^2}{4(4p^2-1)}$. Given the conformal block $G_{\Delta,0}$, from this recursion we can compute $G_{\Delta,1},G_{\Delta,2}, \cdots$ iteratively. We stress however, that the resulting expressions become very messy very soon! 

We are interested in a particular term in the small $v$ limit, namely

\begin{equation}
G_{\Delta,\ell}(u,v) = \log v h_{\Delta,\ell}(u) + \cdots \ .
\end{equation}
Inserting this expansion in the recursion relation above, we obtain a simplified recursion for $h_{\Delta,\ell}(u)$, which can be supplemented with the known expression for $h_{\Delta,0}(u)$ as initial condition. The solutions have a expansion of the form
\begin{equation}
h_{\Delta,\ell}(u)= -\frac{\Gamma(\Delta+\ell)}{\Gamma^2(\frac{\Delta+\ell}{2})} \frac{(d-2)_\ell}{(d/2-1)_\ell} u^{(\Delta-\ell)/2}\left(1+ \sum_{i=1} a^{(i)}_{\Delta,\ell} u^i \right) \ .
\end{equation}
The coefficients $a^{(i)}_{\Delta,\ell}$ have the following structure
\begin{equation}
a^{(i)}_{\Delta,\ell}= \frac{\Gamma(\ell+\alpha-i)\Gamma(1+\alpha-i-\Delta)}{(\ell+2\alpha-\Delta-1)\Gamma(\alpha+1) \Gamma(\alpha+1-\Delta)} p^{(i)}(\Delta,\ell) \ ,
\end{equation}
where $p^{(i)}(\Delta,\ell)$ are polynomials of degree $2i+1$ in both $\ell$ and $\Delta$. The recursion relations for $h_{\Delta,\ell}(u)$ can then be translated into recursion relations for such polynomials:
\begin{eqnarray*}
\frac{\Delta-2 \alpha  -\ell+3}{\Delta-2 \alpha  -\ell+1} p^{(i)}(\Delta,\ell)  + \frac{\alpha +\ell-1}{i-\alpha-\ell+1}p^{(i)}(\Delta+1,\ell-1)= \\
-\frac{(\ell-1) (2 \alpha +\ell-2) (\Delta +\ell-2) (-\alpha +\Delta +i-1)}{4 (\alpha -\Delta ) (\alpha -i+\ell-1)} p^{(i-1)}(\Delta,\ell-2) \\
 -\frac{\Delta  (\ell-1) (2 \alpha -\Delta -1) (\alpha +\ell-1) (2 \alpha +\ell-2) (-2 \alpha +\Delta -\ell+2)^2}{16 (\alpha -\Delta ) (2 \alpha -\Delta +l-5) (2 \alpha -\Delta +\ell-1)} p^{(i-2)}(\Delta+2,\ell-2) \ ,
\end{eqnarray*}
with the initial conditions $p^{(-2)}(\Delta,\ell)=p^{(-1)}(\Delta,\ell) =0$, the known results for $p^{(i)}(\Delta,0)$:
\begin{equation}
p^{(i)}(\Delta,0) = \frac{(-1)^i \left((2 \alpha -\Delta -1) \Gamma (\alpha ) \Gamma \left(i+\frac{\Delta }{2}\right)^2\right)}{\Gamma \left(\frac{\Delta }{2}\right)^2 \Gamma (i+1) \Gamma (\alpha -i)} 
\end{equation}
and enough patience, one can solve for the polynomials $p^{(i)}(\Delta,\ell)$ to any desired order. For instance, we find
\begin{eqnarray}
a^{(1)}_{\Delta,\ell} =\frac{-(d-4) \Delta ^2+\ell^2 (-d+2 \Delta +2)+2 \ell \left(d-\Delta ^2-\Delta -2\right)}{2 (d-2 (\Delta +1)) (d+2 \ell-4)}
\end{eqnarray}
and so on. The coefficients $a^{(2)}_{\Delta,\ell} $ and higher are too complicated to be shown here. 

Finally, for several applications in the body of the paper, we will be interested in the small $u$ limit of the conformal blocks. With some effort, higher order corrections to the collinear conformal blocks can again be computed used the recursion relations above. We find

{\tiny
\begin{eqnarray} \nonumber
f_{\Delta,\ell}(u,v) &=&f^{(0)}_{\Delta,\ell}(v)  + u f^{(1)}_{\Delta,\ell}(v)  +\cdots \ , \\ \nonumber
f^{(0)}_{\Delta,\ell}(v) &=&(1-v)^\ell ~_2F_1\left(\frac{\ell+\Delta}{2},\frac{\ell+\Delta}{2},\ell+\Delta;1-v\right) \ , \\ \nonumber
f^{(1)}_{\Delta,\ell}(v)&=& (-1)^\ell \frac{(\Delta +\ell-1)  (v-1)^{\ell-4} }{(\Delta +\ell-2) (d-2 (\Delta +1)) } \left(8 (\Delta -1) \Delta (v+1) \, _2F_1\left(\frac{1}{2} (\ell+\Delta -4),\frac{1}{2} (\ell+\Delta -2);\ell+\Delta -2;1-v\right)\right.\\ \nonumber
 &-&\frac{4  \left(\Delta ^2 (v (v+6)+1) (d+2 \ell-4)+2 \Delta  \left(\ell \left(-\ell (v-1)^2+(v-10) v+1\right)-4 (d-4) v\right)+(d-2) (\ell-2) \ell (v-1)^2\right)}{(d+2 \ell-4)} \times\\ \nonumber
& &\left. \times \, _2F_1\left(\frac{1}{2} (\Delta +\ell-2),\frac{1}{2} (\Delta +\ell-2);\Delta +\ell-2;1-v\right) \right) \ .
\end{eqnarray}
}

\section{Correction to OPE coefficients}

It is straightforward to apply the algebraic method to the computation of large spin corrections to the OPE coefficients of double trace-like operators. Let us focus in the case of a scalar operator ${\cal O}_\epsilon$ plus its tower of descendants. To linear order in $\delta \hat a = \hat a -1$ and $\gamma$ we obtain
\begin{equation}
\int_0^\infty d j \ K^{\Delta}(j,v) \left( \delta \hat a - \gamma \log(1-v) \right)= -a_\epsilon v^{\Delta_\epsilon/2} \frac{\Gamma(\Delta_\epsilon)}{\Gamma^2(\frac{\Delta_\epsilon}{2})} \left(2\psi(\frac{\Delta_\epsilon}{2})+2\gamma + (\frac{\Delta_\epsilon^2(2\psi(1+\frac{\Delta_\epsilon}{2})+2\gamma )}{4-2d+4\Delta_\epsilon}) v+ \cdots \right) \ ,
\end{equation}
which translates into
{\scriptsize
\begin{eqnarray}\nonumber
\delta \hat a&=& -\frac{c_0}{J^{\Delta_\epsilon}} \left( \psi(\frac{\Delta_\epsilon}{2})+\gamma  \right)\\ \nonumber
& & +\frac{c_0}{J^{\Delta_\epsilon+2}}  \left(\frac{3 (d-2) (\Delta_\epsilon-2 \Delta  +2)^2-\Delta_\epsilon  \left(\psi\left(\frac{\Delta_\epsilon }{2}\right)+\gamma \right) (\Delta_\epsilon  (-12 (\Delta -1) \Delta +\Delta_\epsilon  (\Delta_\epsilon +4)+8)+8)-2 d (\Delta_\epsilon  (-3 \Delta +\Delta_\epsilon +3)+2)}{24 (d-2 (\Delta_\epsilon +1))}\right)\\
& & + \cdots \ ,
\end{eqnarray}
}
where we have used the expansion for the anomalous dimension $\gamma = -c_0/J^{\Delta_\epsilon}+\cdots$. In section five we will be interested in the case $\Delta_\epsilon=2$, $\Delta=\mu-1$, $d=2\mu$, relevant for the critical $O(N)$ sigma model. In this case:
\begin{equation}
\delta \hat a = \frac{(\mu-3)(\mu-1)}{2} \frac{c_0}{J^{\Delta_\epsilon+2}} + \cdots \ .
\end{equation}

\section{$R(s, \mu)$ integral and colored permutations}

Let us consider the integral that appears in the two-loop computation of the $O(N)$ model $R(s, \mu)$, In the large spin expansion it has the structure (92). Both series can be computed exactly 
and we get
\begin{eqnarray}\label{integralR}
R(s, \mu) &=& {\Gamma(\mu - 2)^2 \Gamma({1 \over 2} \sqrt{1 + 4 J_0^2}+ {5 \over 2} - \mu) \over \Gamma({1 \over 2} \sqrt{1 + 4 J_0^2}+ \mu - {3 \over 2})} \\ \nonumber
&+& {\ _3 F_2 (1, {1 \over 2} + {1 \over 2} \sqrt{1 + 4 J_0^2}, 3 - \mu ; {3 \over 2} + {1 \over 2} \sqrt{1 + 4 J_0^2} , {7 \over 2} + {1 \over 2} \sqrt{1 + 4 J_0^2} - \mu ; 1 )
 \over ({1 \over 2} + {1 \over 2} \sqrt{1 + 4 J_0^2}) ({1 \over 2} \sqrt{1 + 4 J_0^2}+ {5 \over 2} - \mu )} 
\end{eqnarray}
where the first line contains all the terms of the type ${1 \over s^{2 \mu - 2 + n}}$ and the second ${1 \over s^{2 + n}}$. We also notice that the second line of (\ref{integralR}) is related to the number
of non-crossing, non-nesting colored permutations ${\rm NCN}_{2,2}(n,k)$ \cite{LYen} (see section 4.3 in that paper). More precisely, we observe the following relation
\begin{eqnarray}
\sum_{n \geq 0} {\rm NCN}_{2,2}(n, \mu - 3) x^{n}= {\ _3 F_2 (1, {1 \over 2} + {1 \over 2} \sqrt{1 + {4 \over x} }, 3 - \mu ; {3 \over 2} + {1 \over 2} \sqrt{1 + {4 \over x} } , {7 \over 2} + {1 \over 2} \sqrt{1 + {4 \over x} } - \mu ; 1 )
 \over x ( {1 \over 2} \sqrt{1 + {4 \over x} } - {1 \over 2} ) ({1 \over 2} \sqrt{1 + {4 \over x} }+ {5 \over 2} - \mu )} .
\end{eqnarray}
We do not understand the origin of this relation. Equivalently, we can write for integer $k$
\begin{eqnarray}
\sum_{n \geq 0} {\rm NCN}_{2,2}(n, k) x^{n}= \sum_{m=0}^{k} {\Gamma(k+1)^2 \over \Gamma(m+1) \Gamma(2 k - m + 2)} {2k + 1 - 2 m \over 1 - (k-m)(1+(k-m)) x} .
\end{eqnarray}

\section{Correlation functions in the $O(N)$ model}

Here we present some explicit formulas for correlation functions in the critical $O(N)$ model. One can rewrite the results of \cite{Lang:1991kp} in the following form
\begin{eqnarray}\nonumber
& &\langle \sigma_a (x_1) \sigma_b (x_2) \sigma_c(x_3) \sigma_d (x_4) \rangle = {\delta_{ab} \delta_{cd} \over \left( x_{12}^2 x_{34}^2 \right)^{\Delta_{\sigma}}} + {\delta_{ac} \delta_{bd} \over \left( x_{13}^2 x_{24}^2 \right)^{\Delta_{\sigma}}}+ {\delta_{ad} \delta_{bc} \over \left( x_{14}^2 x_{23}^2 \right)^{\Delta_{\sigma}}}   \\
&+& {\eta_1 \mu \Gamma(\mu) \over (\mu -2) \Gamma(\mu - 1)^2 N} {f_{a b c d}^{(1)}(u,v) \over \left( x_{12}^2 x_{34}^2 \right)^{\Delta_{\sigma}}} \ , \\ \nonumber
f_{a b c d}^{(1)}(u,v) &=& \delta_{a b} \delta_{c d} \bar D_{1, \mu - 1,1, \mu - 1}(u,v) + \delta_{a d} \delta_{b c} \bar D_{1, \mu - 1, \mu - 1, 1}(u,v) + \delta_{a c} \delta_{b d} \bar D_{\mu - 1, \mu - 1,1,1} (u,v).
\end{eqnarray}
where the definition of the $\bar D$-functions is given for instance in \cite{Dolan:2000ut}. Another correlator that we used in the bulk of the paper to compute some of the relevant three-point couplings is the following
\begin{eqnarray}\nonumber
\langle  \alpha(x_1) \alpha(x_2) \sigma_c(x_3) \sigma_d (x_4)  \rangle &=& {\delta_{c d} \over x_{12}^{2 \Delta_{\alpha}} x_{34}^{2 \Delta_{\sigma}}} \left( 1 + {\Gamma(\mu + 1)\eta_1 \over \Gamma(\mu - 1)^2 N} [\bar D_{\mu - 1 , 2 ,\mu - 2, 1} (u, v) + \bar D_{\mu - 2 , 2 ,\mu - 1, 1} (u, v) ]  \right. \\
& &\left. + {2 \mu (\mu - 1)(2 \mu - 3) \eta_1 \over (\mu - 2)^2\Gamma(\mu - 2) N}  \bar D_{\mu - 1 , 1 ,\mu - 1, 1} (u, v)\right) . 
\end{eqnarray}
Recall that $\eta_1=\frac{2 (2-\mu ) \Gamma (2 \mu -2)}{\Gamma (\mu -1)^2 \Gamma (2-\mu ) \Gamma (\mu +1)}$.

\end{document}